\documentclass[pdflatex,sn-mathphys-num]{sn-jnl}

\usepackage{graphicx}%
\usepackage{multirow}%
\usepackage{amsmath,amssymb,amsfonts}%
\usepackage{amsthm}%
\usepackage{graphicx}
\usepackage{dcolumn}
\usepackage{mathrsfs}%
\usepackage{bm}
\usepackage[title]{appendix}%
\usepackage{float}
\usepackage{xcolor}%
\usepackage{textcomp}%
\usepackage{manyfoot}%
\usepackage{booktabs}%
\usepackage{algorithm}%
\usepackage{algorithmicx}%
\usepackage{algpseudocode}%
\usepackage{listings}%
\usepackage[T1]{fontenc}
\usepackage[utf8]{inputenc}
\usepackage{xurl} 
\usepackage{cancel}
\usepackage[normalem]{ulem}

\usepackage{mathptmx}
\usepackage{dirtytalk}
\usepackage{etoolbox}
\usepackage{subcaption}
\usepackage{lineno}

\theoremstyle{thmstyleone}%
\theoremstyle{thmstyletwo}%

\theoremstyle{thmstylethree}%

\renewcommand{\arraystretch}{1.3}  
\begin{document}

\title[Article Title]{Free oscillations of a standing surface wave and its mechanical analogue}


\author[1]{\fnm{Nikhil} \sur{Yewale}}\email{yewale.nikhil65@gmail.com}

\author[2]{\fnm{Sakir} \sur{Amiroudine}}\email{sakir.amiroudine@u-bordeaux.fr}

\author*[1]{\fnm{Ratul} \sur{Dasgupta}}\email{dasgupta.ratul@iitb.ac.in}

\affil*[1]{\orgdiv{Chemical Engineering Department}, \orgname{Indian Institute of Technology Bombay}, \orgaddress{ \city{Powai, Mumbai}, \postcode{400076}, \state{Maharashtra}, \country{India}}}

\affil[2]{\orgname{Univ. Bordeaux, CNRS, Bordeaux INP, I2M, UMR 5295, F-33400}, \orgaddress{\city{Talence}, \country{France}}}


\abstract{	We present an analogy between natural oscillations of the standing wave type on a pool of liquid with an
 interface and a mechanical oscillator model. It is shown that the equations of motion governing both systems
  have qualitatively similar solutions - trivial as well as time-periodic with finite amplitude. The time-periodic solutions can be linearly unstable in both cases depending on the oscillation amplitude, thereby leading to interesting dynamics. Linear stability results of both systems are discussed in detail; a novel Mathieu-like equation is derived for the stability of the standing wave to a super-harmonic perturbation. \textcolor{black}{This is obtained through a much simpler        
  	approach that yields linear stability results while also reinforcing the analogy.} Analytical predictions are compared against numerical solutions to the full nonlinear governing equations for both systems. A good match is obtained in most cases with theory; mismatches are further analysed and the limitations of this analogy are also pointed out.}




\maketitle

\section{Introduction}\label{sec1}

Stability analysis of vibrating systems, whether discrete (finite degrees of freedom) or continuous (infinite degrees of freedom), have been of perpetual interest to engineers. This has become possible due to major strides made over a century in perturbation techniques \cite{bender2013advanced} as well as progress in computational algorithms. The classical textbooks by Stoker \cite{Stoker} and Nayfeh \cite{Nayfeh} present the stability of a range of vibrating systems, of interest to mechanical and electrical engineers. Mathematical techniques for modeling finite degrees of freedom systems also extend to the study of vibration phenomena in continuum mechanics, both fluid as well as solid. In interfacial fluid mechanics, standing surface waves have been of recurrent interest due to their relevance to the phenomena of sloshing \cite{ibrahim2005liquid}. Sloshing occurs quite commonly in seemingly varied situations such as  partly filled liquid tanks subject to erratic horizontal accelerations \cite{turner2024dynamic} or sea surface oscillations inside ``moonpools" in offshore oil production barges \cite{molin2001piston,miles2002gravity,chu2024nonlinear}, to name only a few among myriads of such examples. \textcolor{black}{Our aim here is to illustrate an analogy between natural oscillations of a mechanical vibrating system and its continuous analogue i.e. an interfacial wave. A similar analogy is well-known, albeit for forced oscillations. This is the Kapitza pendulum which has been used as a mechanical analogue to understand the Faraday instability on the surface of a vertically vibrated pool of liquid; in \citet{rajchenbach2015faraday}, the authors write insightfully:}

\textit{\textcolor{black}{	
\say{Faraday waves are often analysed in analogy with the parametric excitation
	of a pendulum \cite{fauve1998pattern}. Although this simple model presents an obvious
	pedagogical interest, in particular to introduce the Mathieu equation and to show
	that the first resonance corresponds to half of the forcing frequency, use of the
	parametric pendulum analogy is often misleading. Indeed, a major difference is that
	the eigenfrequency of a freely oscillating pendulum is unique, whereas free unforced
	water waves exhibit a continuous spectrum of mode frequencies.}
}
}

\textcolor{black}{
In this study, we develop a similar analogy, but between \textit{natural} oscillations of a fluid surface and that of a mechanical oscillator. We show that in addition to several qualitative similarities, there remain important differences between the two systems. Our study analyses these in detail.}
\subsection{\textcolor{black}{Literature review}}\label{lit}
\textcolor{black}{In what follows, we briefly summarise what is already known about the mechanical and interfacial oscillator. We begin with the latter first.}
\subsubsection{\textcolor{black}{Finite amplitude  waves (interfacial oscillator)}}
Our interest here is concerned with \textit{finite amplitude}, standing waves and their stability. Such waves often appear on an air-water interface, originating from an initial interface distortion,  \textcolor{black}{without} any external forcing except for what is applied initially. Apart from fundamental relevance, these natural oscillations are often also of engineering significance. For instance, a problem of interest to coastal engineers designing circular harbours connected to the open ocean, is to predict the natural frequencies of the ocean surface contained within the circular ocean basin, in order to preclude the occurrence of resonant seiches \cite{CoastalWiki,mack1962periodic, miles1974harbor}. Computing the shape of the ocean surface for \textit{finite-amplitude}, time-periodic oscillations (with gravitational restoring force), even for such a common geometry as a cylindrical basin is an arduous task, requiring lengthy analytical calculations \cite{mack1962periodic}. Further complexities arise from the fact that such finite amplitude, natural oscillations, are often unstable at large oscillation amplitude, leading to aperiodic behaviour and distortion of the interface shape with complicated accompanying wave dynamics. Not surprisingly, analyzing these instabilities requires significant numerical  \cite{zhu2003three} or analytical effort towards accurately obtaining the nonlinear, time-periodic, base-state  \cite{concus1962standing,saffman1979note, schwartz1981semi,rottman1982steep,rycroft2013computation}, whose linear stability is then sought \cite{Mercer1992,zhu2003three}. This is done either through Floquet analysis \cite{Mercer1992}, \textcolor{black}{Bloch analysis \cite{Cross_Greenside_2009}} or alternatively employing weakly nonlinear equations such as the Zakharov equation  \cite{okamura1984instabilities}. \textcolor{black}{Inspired by the analogy with the mechanical oscillator, we will show in section \ref{sec:stabilityInterfacial} that a simplified Mathieu-like equation, captures linear stability properties qualitatively, of the standing wave.}
\subsubsection{\textcolor{black}{Mechanical oscillator, their analogy with continuum systems and literature survey}}
While the importance of the intensely mathematical approaches, \textcolor{black}{such as Floquet or Bloch analysis mentioned above}   can hardly be overstated, for complementary physical understanding, it seems useful to seek toy models with finite degrees of freedom (implying that the governing equations are ordinary instead of partial, in the toy model). These equations admit analogous, finite-amplitude oscillations and parameter regimes where such oscillations may be potentially unstable. These models are typically much easier to analyze mathematically and via analogy, permit an intuitive understanding of more complicated systems such as interfacial natural oscillations described earlier, while also delineating possible differences. \textcolor{black}{An example based on Faraday waves has been discussed above. Similar analogies also extend to stratified fluids, see for example \citet{koszalka2005vibrating}}.

In the following, we first discuss oscillations of a fluid interface and follow it up with presentation of natural oscillations of a spring-mass system \textcolor{black}{first described by \citet{YANG1967} and later by \citet{KovacicRand}. \citet{YANG1967} discussed the vibrations of a particle on a plane for natural \cite{YANG1967} and forced oscillations \cite{YANG1968}. These studies discussed normal-modes of vibrations for both cases. Hill and subsequent Mathieu equations were derived for studying the stability of the equilibrium configurations. However, the stability charts  in these studies were all based on the Mathieu equation. Further studies such as \cite{RandandFu}, \cite{KovacicRand} confirmed the stability of this system using a perturbative approach employing the Hill equation. Related studies \cite{PECELLI198057} have also been conducted on a non-elastic spring-mass system with the same geometric configuration. To our knowledge, a complete stability chart based on the Hill equation, and numerical validation through time-evolution of the full nonlinear system has not been reported before and is done here. We show that there are qualitative similarities between stability properties of the mechanical oscillator and the interfacial oscillator; this thus extends the Faraday wave-pendulum analogy discussed in the introduction, to natural oscillations.}

\subsection{\textcolor{black}{Outline of the study}}
 \textcolor{black}{Our study is organized as follows: Section \ref{finiteamptheoryandsims} describes free oscillations of a finite-amplitude, surface-gravity wave using numerical simulations of the incompressible Euler's equations and compares it against theory. Section \ref{sec2} describes the equations of motion for the mechanical oscillator analogue. Section \ref{sec:LinearEquilibrium} discusses the linear stability analysis for trivial and finite-amplitude, time-periodic solution to the equations of motion of the oscillator; in particular, a novel stability chart based on the Hill equation is presented in this section \ref{sec:hilleqn}. These predictions are then validated against numerical solutions to the full nonlinear equations of motion in section \ref{sec:nonlinearevolution}. The analogy, both for trivial and non-trivial solutions, of the mechanical and the interfacial oscillator are established in sections \ref{sec:trivial} and \ref{sec:nontrivial} respectively. In section \ref{sec:stabilityInterfacial}, we present a reduced order model which leads to a novel Mathieu-like equation governing the stability of the interfacial wave to a super-harmonic perturbation; this analysis further reinforces the analogy. We conclude by discussing the similarities and differences between the two systems. Table \ref{tab:table1} summarizes the novel contributions, section wise, in our present work.
{\color{black}
\begin{table}[h!]
	\centering
	\renewcommand{\arraystretch}{1.7}
	\setlength{\tabcolsep}{6pt}
	\begin{tabular}{>{\color{black}}c @{\hspace{1cm}} >{\color{black}\raggedright\arraybackslash}m{0.7\textwidth}}
		\hline
		\textbf{Section} & \textbf{Novel contributions} \\
		\hline
		 \ref{sec:hilleqn} &  Stability chart for mechanical oscillator based on Hill equation \textcolor{black}{(fig. \ref{fig:hill_stab})} \\
		 \ref{sec:nonlinearevolution} & Evolution of full nonlinear equations, Hill equation and Mathieu equation for parameters in stable and unstable regime (numerical validation of linearised predictions) \textcolor{black}{(figs. \ref{fig:numericalSol_Timetrace} and \ref{fig:comp})} \\
		 \ref{sec:trivial}, \ref{sec:nontrivial} &  Establishing the analogy of trivial and non-trivial solutions for interfacial oscillator with the mechanical oscillator. \\
		 \ref{sec:stabilityInterfacial} & Low-order representation of stability analysis for a weakly non-linear interfacial oscillator \textcolor{black}{(Mathieu-like eqn. (\ref{eq33}))}\\
	\end{tabular}
	\caption{Sectional summary and novel contributions}
	\label{tab:table1}
\end{table}}
}
\section{\textcolor{black}{Problem Descriptions and Methodology}}
\subsection{\textcolor{black}{Interfacial oscillator - Time periodic, standing waves of small and large amplitude on a liquid pool }}\label{finiteamptheoryandsims}
\begin{figure}[htbp]
	\centering
	\includegraphics[scale=0.3,trim={8cm 0 0 2cm},clip]{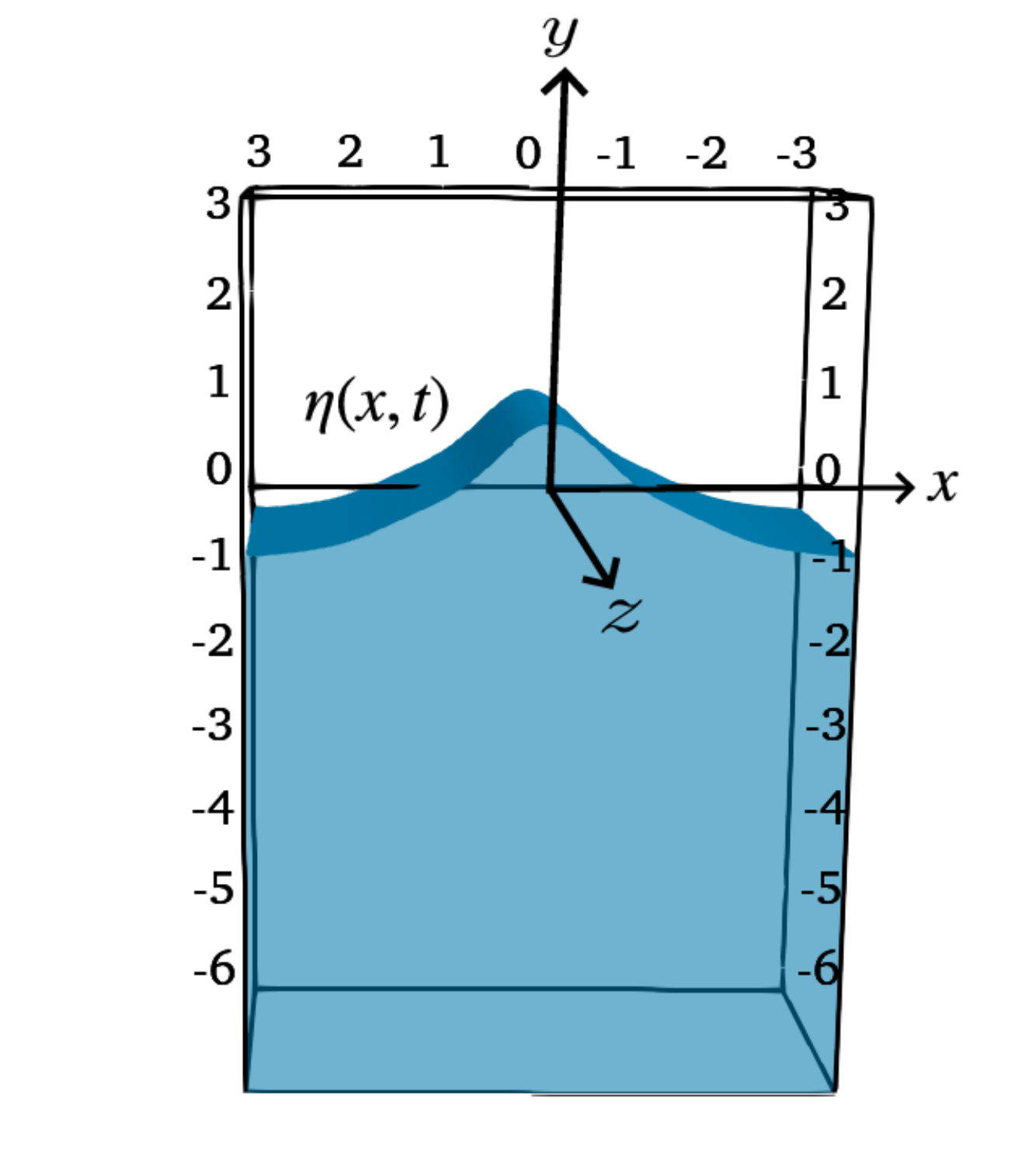}
	\captionsetup{width=\columnwidth}
	\caption{ An interface between air and water initialized as a standing wave at time $t=0$ using the $\mathcal{O}(\hat{A}^5)$ formula in eqn. \ref{eq2} taken from \citet{Penney1952}. The perturbed interface $\eta(x,0)$ is depicted for the value $\hat{A}=0.592$, which is high steepness and thus the crest of the deformation is quite sharp while the trough is flat, when compared to the cosine wave of eqn. \ref{eq1}. The wavelength of the initial interface is $2\pi$ meters . The density of water is set to $1000\, kg/m^{3}$ and that of the upper fluid is $1\,  kg/m^{3}$. Surface-tension ($T$) is set to $0.072\, N/m$ and $g=9.81\, m/s^{2}$. The origin of the vertical ($y$) axis lies at the undisturbed interface ($y=0$). In the simulation the $z$ direction is absent.}
	\label{fig0_SurfaceWave}
\end{figure}
In this section, we commence our analysis with a description of finite amplitude, standing waves \textcolor{black}{(natural oscillations)} on an air-water interface. To facilitate further discussion on the proposed analogy, we will refer to the oscillations studied in this section as that of an `interfacial oscillator' . Fig. \ref{fig0_SurfaceWave} represents a schematic of a pool of water (blue) with air on top (white). We simulate natural oscillations of the standing wave type on the gas-liquid interface, in two dimensions. Oscillations are generated by imposing an initial interfacial deformation at $t=0$ (see below for the precise form of the deformation) and we integrate the equations of motion (incompressible Euler's equations with gravity and surface-tension) in time using the open-source code Basilisk \cite{popinet-basilisk}. The restoring force for the interfacial oscillation is primarily gravity with a small contribution from surface tension. This is ensured by restricting the length scale (wavelength) of our initial interfacial deformation to be $2\pi$ meters, significantly greater than the air-water capillary length scale $\left(l_c  = \sqrt{\dfrac{T}{\rho g}} = \sqrt{\dfrac{0.072}{1000 \times 9.81}} = 2.72 \times 10^{-3}\, m \approx 2.72\, mm\right)$, 
thus ensuring that we are simulating surface gravity waves with negligible capillary effects. Due to the relatively low kinematic viscosity of water and the large wavelengths under consideration (implying high Reynolds number), we \textcolor{black}{consider inviscid flow} \textcolor{black}{and thus neglect the momentum boundary layer(s) that would otherwise be generated.} Thus in our simulations and theory later on, the boundary layers at the interface and at solid boundaries are ignored in a first approximation. \textcolor{black}{We employ an adaptive grid in Basilisk with maximum grid level $10$.} Our simulations have been checked for grid convergence by running these at a higher resolution of level $11$ as well, although we only report results from level $10$ here. Free-slip, and no-penetration boundary conditions are imposed on all computational domain boundaries throughout the course of the simulation. The contact angle is maintained at $\pi/2$ in the simulation. Note that our simulations are two-dimensional although fig. \ref{fig0_SurfaceWave} is a three dimensional rendition, for ease of visualization. 
\begin{figure}
	\centering
	\makebox[\textwidth][c]{\includegraphics[scale=0.38,trim={0 0 0 0}]{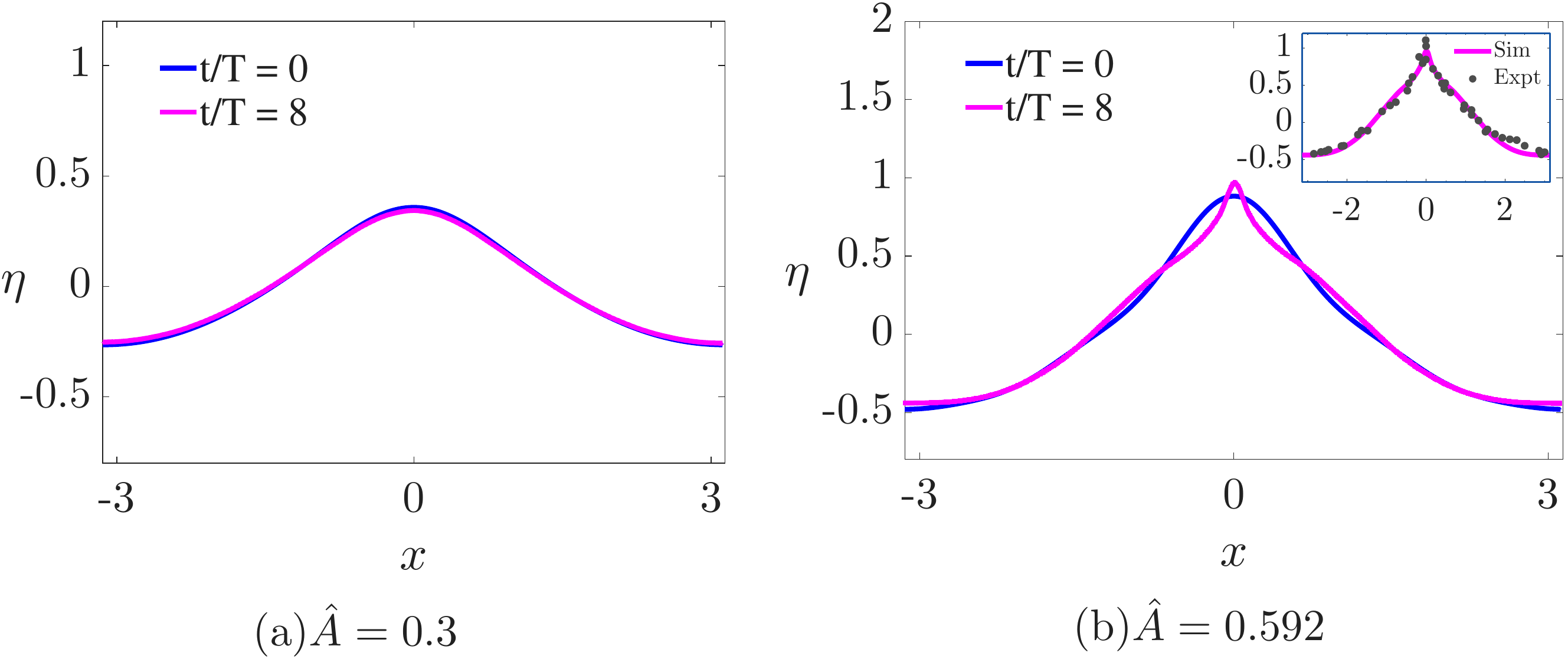}}
	\caption{Panel (a) Comparison between the analytical solution of \citet{Penney1952} and numerical simulation for $\hat{A}=0.3$ after eight time periods $T$, see formula for $T$ below eqn. \ref{eq2} (b) Comparison for $\hat{A}=0.592$. In this panel, note the mismatch around the crest of the wave and the formation of a sharp corner in the simulated wave profile. In the inset, comparison of our profile (solid line) with a similar sharp crested profile seen in the experimental results of \citet{Taylor1953} (dots) is provided, when their standing wave \textcolor{black}{is} reaching its maximum height.} 
	\label{fig:standing_waves_0_2T}
\end{figure}

Before presenting results, it is useful to recall what may be intuitively expected from linearized potential flow theory. When the amplitude of interfacial deformation is small (compared to  the wavelength), one expects linear behaviour in time. For example, if the interface were to be deformed from its flat state at $t=0$ as \textcolor{black}{:} 
\begin{align}
	\eta(x,t=0) = a_1\cos(kx) \label{eq1}
\end{align}
with zero fluid velocity everywhere ($k=2\pi/\lambda$, $\lambda$ being the wavelength), we anticipate that a standing wave of frequency $\sqrt{gk}$ will result when $a_1 << 2\pi/k$. This is the deep-water limit ($H\rightarrow\infty$, $H$ being the undisturbed water depth) of the natural frequency of a linearized surface-gravity wave  on a layer of liquid of depth $H$ viz.  $\omega_0 = \sqrt{gk\tanh(kH)}$ (see Chapter 8 in book by  \citet{Kundu2016} ). This deep water approximation to the natural frequency is valid, when the perturbation wavelength $\lambda = \dfrac{2\pi}{k} << H$, this being ensured in our simulation. It is perhaps natural to ask, if the standing waves expected in the linear regime, may continue to be obtained as $a_1$ is increased relative to the wavelength of the cosine in eqn. (\ref{eq1}), in our simulations. 

The answer to this was first obtained by Rayleigh \cite{strutt1915deep} who demonstrated that one can obtain time-periodic, standing waves for a range of values of the non-dimensional parameter $\hat{A} \equiv a_1k$ ($0 < \hat{A} \leq \hat{A}_c$ where $\hat{A}_c \approx 0.6202$ \cite{Mercer1992}). However, the interface shape has the form of eqn. (\ref{eq1}) only when $a_1$ is sufficiently small. This was demonstrated by solving the nonlinear, potential flow equations in a perturbative manner using $\hat{A}$ as a small parameter. Here onwards, we will refer to $\hat{A}$ as (wave) steepness following standard terminology. \textcolor{black}{Rayleigh} demonstrated that for each value of $\hat{A}$ (in the aforementioned range), the interface adopts a particular shape which generates time-periodic oscillation. The time period also depends on the wave steepness $\hat{A}$ and this can be obtained analytically\textcolor{black}{.} \textcolor{black}{Rayleigh} reported his analysis for a liquid with a free surface (i.e. by setting the density of air to zero and neglecting surface tension) while assuming spatial periodicity in the horizontal coordinate. The interface shape is thus expressed as a Fourier series in $x$, involving integer harmonics of the wavenumber $k$ in eqn. (\ref{eq1}). Importantly, when $\hat{A} << 1$, the interface shape in Rayleigh's formula reduces to a cosine of wave-number $k$ and amplitude $a_1$. For larger values of $\hat{A}$, there are however terms containing harmonics of $k$, with the right hand side of eqn. (\ref{eq1}) being the leading order term in this infinite series representation.
Since this seminal work on the form of a nonlinear standing wave, the task of determining the ``shape'' of the finite-amplitude standing wave as a function of $\hat{A}$ has been revisited by several authors over the last seventy-five years commencing with the work of \citet{Penney1952} who \textcolor{black}{have considered the base-state} up to $\mathcal{O}(\hat{A}^5)$ \cite{Penney1952,concus1962standing, saffman1979note,schwartz1981semi}. 

In fig. \ref{fig0_SurfaceWave}, we depict the initial condition that is used to initiate the interface in our numerical simulations \cite{popinet-basilisk}. This is obtained from the $\mathcal{O}(\hat{A}^5)$ formulae obtained by \citet{Penney1952} for time-periodic motion. The deformed interface may be compactly represented at \textcolor{black}{$t=(\pi/(2\omega_0))$} as (see below for definition of $\omega_0$): 
\begin{eqnarray}
	k\eta(x,0) = \tilde{\eta} = \dfrac{b_0}{2} + \sum_{n=1}^{5} b_n \cos (nkx), \label{eq2}
\end{eqnarray}
where expressions for $b_0, b_1,\ldots b_5$ are \textcolor{black}{ expressed in appendix \ref{app4}}. In fig. \ref{fig:standing_waves_0_2T}, we depict two simulation results. In both panels, the simulation has been continued upto eight time periods $T$ ($k=1$ in both simulations) where :

$T\equiv \dfrac{2\pi}{\omega_0}=\dfrac{2\pi}{\sqrt{gk}\left( 1 - \dfrac{\hat{A}^2}{4} - \dfrac{13}{128} \hat{A}^4\right)^{1/2}}$\cite{Penney1952}. Note that in the limit of $\hat{A}\rightarrow 0$, $T \approx \dfrac{2\pi}{\sqrt{gk}}$ is just the time-period of a linearized standing wave. \textcolor{black}{This} fifth order analytical expression for $\eta(x,t)$ also serves to benchmark the accuracy of our numerical simulation. 

An obvious difference is seen in the panels (a) and (b) of fig. \ref{fig:standing_waves_0_2T}. For $\hat{A}=0.3$, there is hardly any difference between the analytical prediction and the numerical computation after eight time-periods i.e. at $t=8T$. For $\hat{A}=0.592$ (panel (b) of the same figure) however, there is a significant difference. In particular, note the sharpening of the crest of the standing wave. This difference at $\hat{A}=0.592$, between the numerically evolved and exact solution was also noted by \citet{saffman1979note}, see their fig. $4$. In fig. \ref{fig:standing_waves_0_2T}, panel (b) inset, we compare the interface profile obtained from our numerical solution to the experimental results of \citet{Taylor1953}, both profiles displaying sharp crests. Interestingly, while such pointed crests were also seen in the numerical simulations by \citet{saffman1979note}, they did not comment on the origin of this. Why is there a difference between the simulations and the analytical expression in panel (b) of fig. \ref{fig:standing_waves_0_2T} at $t=8T$ but not in panel (a) where $\hat{A}$ is much smaller? Is there any possible instability at large $\hat{A}$ which could cause this ? Computational inaccuracies seem to be an unlikely candidate to explain these because despite significantly different algorithms for solving the potential flow equations, these seem to appear in the simulations of \citet{saffman1979note} as well as \textcolor{black}{in the present study}. We will return to this question at the end of \textcolor{black}{section 3}, when we discuss the stability of the time-periodic solution.

\textcolor{black}{In the following, we summarise the equations governing the motion of the interfacial oscillator, making reasonable approximations}. Consistent with the inviscid approximation discussed earlier, we assume potential flow implying that the perturbation velocities are derived from a potential i.e. $\bm{u}=\bm{\nabla}\Phi$. The equations governing the motion are standard (for simplicity we take the deep water approximation and neglect surface tension in our analytical model, although this is present in the simulations). Refer to fig. \ref{fig0_SurfaceWave}, the air-water interface and the dynamics in the water layer is governed by the Laplace equation along with no-penetration conditions at the side-walls ($y=\pm\pi/k$ in addition to boundedness conditions for $y\rightarrow-\infty$). Neglecting any pressure fluctuations in air and setting the air density to zero, the equations governing the motion of the bulk liquid and the interface along with boundary conditions are:
\begin{align}
	&\dfrac{\partial^2 \Phi}{\partial x^2} + \dfrac{\partial^2 \Phi}{\partial y^2} = 0, \label{eq12}\\
	&\dfrac{\partial \Phi}{\partial x} = 0 \quad \text{at } x = \pm \dfrac{\pi}{k}, \label{eq13}\\
	&\dfrac{\partial \Phi}{\partial y} \to 0 \quad \text{as } y \to -\infty \text{ for all } t. \label{eq14}
\end{align}
As we neglect the density of air (compared to water in our analytical model), henceforth we refer to the interface as a `free-surface' (stress free). This free-surface is mathematically defined as $y=y_s(x,t)$ while $\Phi(x,y,t)$ is the velocity potential in the liquid (water). In the following we have set the air pressure to zero and neglected the pressure jump at the free-surface consistent with the neglect of surface-tension in our model. Kinematic and dynamic boundary conditions need to be imposed at the free-surface and these boundary conditions supplement eqns. (\ref{eq12})-(\ref{eq14}). These boundary conditions are: 
\begin{align}
	&\dfrac{\partial y_s}{\partial t} + \left(\dfrac{\partial\Phi}{\partial x}\right)_{y=y_s}\left(\dfrac{\partial y_s}{\partial x}\right) = \left(\dfrac{\partial\Phi}{\partial y}\right)_{y=y_s}  \label{eq15}  \\
	&\left(\dfrac{\partial \Phi}{\partial t}\right)_{y=y_s} + gy_s + \left(\dfrac{1}{2}(|\bm{\nabla}\Phi|^2)\right)_{y=y_s} = 0 \label{eq16} 
\end{align}
By construction, the above equations are applicable only at the free-surface $y=y_s(x,t)$. Physically speaking, eqn. (\ref{eq15}) represents the \textcolor{black}{free-surface as material  surface} while eqn. (\ref{eq16}) is the Bernoulli equation applied at the free-surface.
\subsection{The analogue of an interfacial oscillator - a mechanical oscillator\label{sec2}}
\begin{figure}[h]
	\centering
	\includegraphics[scale=0.7,trim={0 0 0 0}]{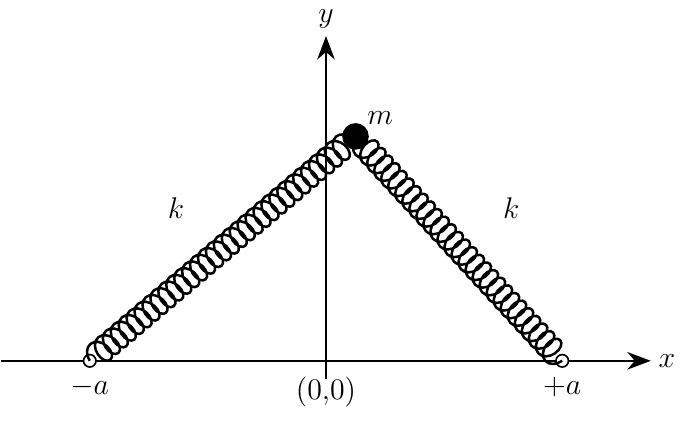}
	\caption{A spring mass system originally studied in \citet{YANG1967} and explored pedagogically in \citet{Rand}. The (identical) linear springs are pivoted at $x=\pm a$ ($a>0$) and have rest lengths $L$ and spring constants $k$. The point mass can move freely on the $x-y$ plane and its motion is restored by the two springs.}
	\label{fig1_springmass}
\end{figure}
In this section, we present an analogue mechanical oscillator which exhibits several qualitative similarities to the interfacial oscillator discussed previously. While this example is not new (see \citet{YANG1967}, \citet{KovacicRand} and \citet{Rand}), we explore this model numerically further than these prior studies. More importantly, the analogy established here between the spring-mass system and the fluid system with an interface has not been presented before, to our knowledge.

Figure \ref{fig1_springmass} depicts a spring-mass system with two degrees of freedom, implying that the mass can move along $x$ and $y$ axes simultaneously. The springs are linear with spring constant $k$ and rest length $L$ and the ends of these springs are attached at points $x=\pm a$ \cite{Rand}. In its equilibrium state, the (point) mass $m$ remains at the origin ($x=y=0$) of the plane and we further assume both springs to be in tension in this state i.e. $a > L$. In Appendix A, it is shown that equations governing $x(t)$ and $y(t)$ are (overdots indicate differentiation with respect to time):
\begin{align}
	&m\bm{\ddot}{x}(t) + k \left( 1 - \dfrac{L}{\sqrt{(a+x(t))^2 + y(t)^2}} \right)(a+x(t)) - k\left(1- \dfrac{L}{\sqrt{(a-x(t))^2 + y(t)^2}}  \right)(a-x(t))=0 \label{eq3}\\
	& \quad m\bm{\ddot}{y}(t) + k \left( 1 - \dfrac{L}{\sqrt{(a+x(t))^2 + y(t)^2}} \right)y(t)  + k\left(1 - \dfrac{L}{\sqrt{(a-x(t))^2 + y(t)^2}}  \right)y(t)=0.  \label{eq4}
\end{align}
Equations (\ref{eq3}) and (\ref{eq4}), are coupled, \textit{non-linear} (kinematic nonlinearity, see discussion in the introduction of \citet{YANG1967}), ordinary differential equations governing the two-dimensional motion of the mass $m$. Note that these equations admit the trivial solution $x(t) = y(t)=0$ which physically corresponds to the mass $m$ being at the origin of the figure \ref{fig1_springmass} at all time - this is an equilibrium configuration as the net force on the mass is zero in this configuration. \textcolor{black}{In the context of our stated aims, our task is now to describe analogies between the set of nonlinear ODE's in eqns. \ref{eq3},\ref{eq4} and the nonlinear PDEs in eqns. (\ref{eq12}-\ref{eq16}). Observe that both sets admit the trivial solution with $x(t)=y(t) = 0$ (particle at the origin) in the former and $y_s(x,t) = \Phi(x,y,t) = 0$ (flat interface with quiescent fluid) in the latter. In addition, both sets also admit non-trivial exact solutions (discussed later). We examine the linear stability of these solutions next, focussing on the mechanical oscillator. In a similar spirit, we will return to the linear stability of exact solutions for the interfacial oscillator equations, subsequently in section \ref{sec:interfacialoscillator}.}
\section{Results}\label{sec:results}
\subsection{\label{sec:LinearEquilibrium}  Linear stability of trivial and non-trivial solutions:}
\textcolor{black}{Consider the trivial solution $x(t)=y(t)=0$ for equations \ref{eq3},\ref{eq4}}. Employing the standard decomposition $x(t) = 0 + \delta x(t)$ and $y(t) = 0 + \delta y(t)$ \cite{Rand} in eqns. (\ref{eq3}) and (\ref{eq4}), and retaining up-to linear terms in $\delta x(t)$ and $\delta y(t)$, we obtain (see \textcolor{black}{Appendix} \ref{app1} for algebra) :
\begin{align}
	&\delta\bm{\ddot}{ x} + \omega_0^2\delta x = 0 \label{eq5}\\
	\text{and}\quad&\delta\bm{\ddot}{ y}  + \omega_0^2 \left(1 - \dfrac{L}{a}\right)\delta y = 0, \label{eq6}
\end{align}
where $\omega_0^2 \equiv \left(\dfrac{2k}{m}\right)$. For $L < a$ (as assumed earlier), both equations (\ref{eq5}) and (\ref{eq6}) can be readily solved using linear combinations of $\cos(\omega_0t)$ and $\sin(\omega_0t)$ and thus predict oscillatory time-periodic behaviour of the perturbations, and no growth. 

The more interesting case is to look for finite amplitude, time periodic solutions to eqns. (\ref{eq3}) and (\ref{eq4}). As is well-known \cite{Rand}, in addition to the trivial solution discussed earlier, $x_b(t)=A\cos(\omega_0t), y_b(t)=0$ is also an exact solution of the nonlinear differential eqns. (\ref{eq3}) and (\ref{eq4}), for any value of $A$ ($|A/a|<1$ as discussed later). That a periodic motion with finite amplitude might be possible and can also be intuitively inferred from the configuration of the mass in fig. \ref{fig1_springmass}. Like earlier, we now consider the stability of this exact solution. An important difference with the previous analysis is that, in the present case the base-state whose linear stability is sought is \textit{time-periodic}. Consequently, the equations governing the perturbations will not be equations with constant coefficients but in general, those with time-periodic coefficients. Such equations are well known and can have unstable solutions, that we seek. \textcolor{black}{The principal novel results in the linear stability analysis that follows, is the demonstration that the Hill equation (section \ref{sec:hilleqn}) provides a much better temporal description of the instability compared to the Mathieu equation studied earlier in this context \cite{Rand}.}
\subsubsection{Hill equation}\label{sec:hilleqn}
\begin{figure}
	\centering
	\includegraphics[scale=0.55]{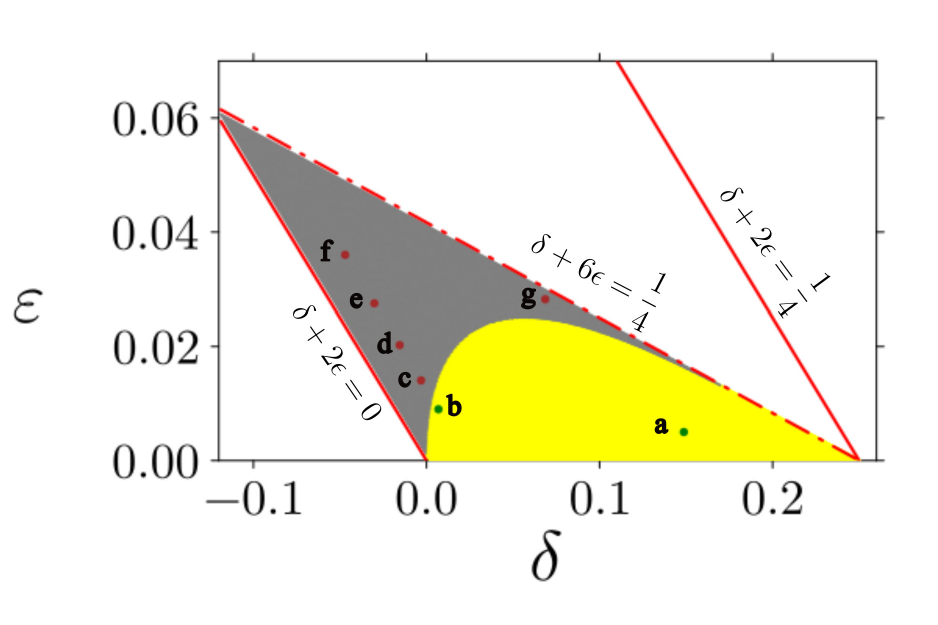}
	\captionsetup{width=\columnwidth}
	\caption{Stability chart of the Hill equation (\ref{eq8}) on the $\epsilon-\delta$ plane. The two lines are given by the formula $\delta + 2\epsilon = 1/4$ and $\delta+2\epsilon=0$. Yellow region - Stable, Grey region - unstable. This chart was generated via Floquet analysis on the $(L/a)-(A/a)$ space with constraints $0<(L/a)<1$ and $|A/a| < 1$. These are then converted into charts in the $\delta-\epsilon$ space using eqns. \ref{eq10}. Note that the white space between the dash-dotted line ($\delta+6\epsilon=1/4$) and the solid line ($\delta + 2\epsilon=1/4$), corresponds to the somewhat unphysical situation $|(A/a)| > 1$ and is not permitted. The dash-dotted line  corresponds to $|A/a|=1$. The points (a), (b),...,(f) correspond to the trajectories represented in fig \ref{fig:numericalSol_Timetrace}.} 
	\label{fig:hill_stab}
\end{figure}

By perturbing the base-state $x(t)= x_b(t) + u(t) = A \cos(\omega_0 t) + u(t), y(t)=y_b(t) + v(t) = 0 + v(t)$ ($x_b(t)=A\cos(\omega_0t), y_b(t)=0$) and substituting in eqns. (\ref{eq3}) and (\ref{eq4}), we obtain the following linearised equations governing the perturbations $u(t)$ and $v(t)$\cite{Rand}:
\begin{align}
	&\ddot{u} + \omega_0^2\;u(t) =0, \label{eq7} \\
	&\ddot{v} + \dfrac{2k}{m} \left[1-  \dfrac{\left(\dfrac{L}{a}\right)}{1 - \left(\dfrac{A}{a}\right)^2\cos^2\left(\omega_0t\right)} \right]v(t) = 0. \label{eq8}
\end{align}
The equation for $u(t)$ is simple and its solution implies that perturbing the oscillatory base-state solution horizontally, does not render it unstable. Eqn. (\ref{eq8}) for $v(t)$ is more interesting, as its coefficient (the term inside square brackets) is time-periodic. Importantly, this time-periodic term contains several harmonics of $2\omega_0t$. To see this, we define the expression inside the square brackets in eqn. (\ref{eq8}) as $f(t)$. For $\omega_0 = 1,\; L/a = 0.9,\; A/a = 0.5$, the Fourier series of $f(t)$ is obtained from Mathematica as \cite{Mathematica}:
\begin{align}
	&&f(t) \approx -0.0392303 + 2.4 \times 10^{-7} \cos(t) - 0.149227 \cos(2t) + 2.4 \times 10^{-7} \cos(3t)\nonumber \\ 
	&& - 0.0107138 \cos(4t) + 2.4 \times 10^{-7} \cos(5t) - 0.000768991 \cos(6t) + \ldots, \nonumber
\end{align}
where the presence of several harmonics of $2\omega_0$ is apparent, notably the even harmonics being dominant. Eqn. (\ref{eq8}) is the well-known Hill differential equation \cite{magnus2013hill} which can have stable as well as unstable solutions. As explained in \textcolor{black}{Appendix}  \ref{Floquet}, we undertake Floquet analysis of eqn. (\ref{eq8}) (\citet{bender2013advanced}) and this generates the stability charts in the non-dimensional parameter space of $\delta-\epsilon$, as shown in fig. \ref{fig:hill_stab}. The definitions of these parameters are provided in expressions (\ref{eq10}) in terms of $(L/a)$ and $(A/a)$. 

We constrain ourselves to the physically meaningful range $0 < \left(L/a\right) < 1$ and $|A/a| < 1$, refer fig. \ref{fig1_springmass}. The first inequality may be used in the definition of $\delta$ and $\epsilon$ in eqns. (\ref{eq10}) to obtain $\delta + 2\epsilon - 1/4 < 0$ and $\delta + 2\epsilon > 0$. The region between the parallel straight lines in fig. \ref{fig:hill_stab} satisfy these two inequalities. Thus by definition, we can only choose those values of $\delta$ and $\epsilon$ which lie between these straight lines. Further, in fig. \ref{fig:hill_stab}, the white region in between the two parallel lines is also prohibited due to the inequality $|A/a| <1$ which translates to $\delta + 6\epsilon < 1/4$, see caption of the figure. The accessible region in this stability chart of the Hill equation (\ref{eq8}) as indicated in fig. \ref{fig:hill_stab}, thus shows two regions in grey and yellow. The region in yellow is stable while that in grey is predicted to be unstable. 

\subsubsection{Mathieu equation}
As shown in \citet{Rand}, eqn. (\ref{eq8}) may be further simplified into the well-known Mathieu equation (\ref{eq9}), by retaining only terms upto $O(A^{2})$ in eqn. (\ref{eq8}). 

\begin{figure}[H]
	\centering
	\begin{subfigure}{0.8\columnwidth}
		\centering
		\includegraphics[scale=0.5,trim={0 0 0 0},clip]{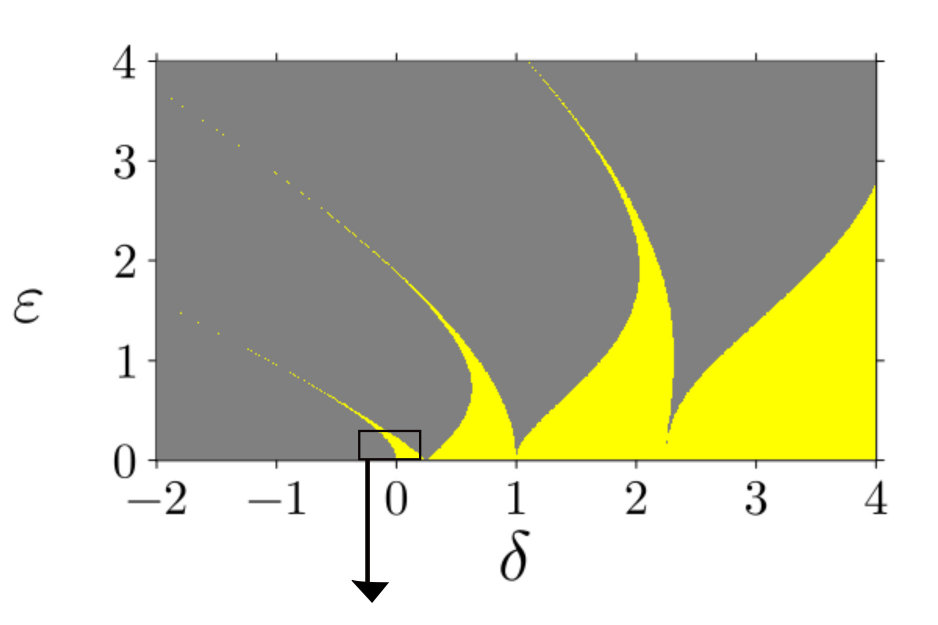}
		\caption{}
	\end{subfigure}
	\vfill
	\begin{subfigure}{0.8\columnwidth}
		\centering
		\includegraphics[scale=0.5,trim={0 0 0 0},clip]{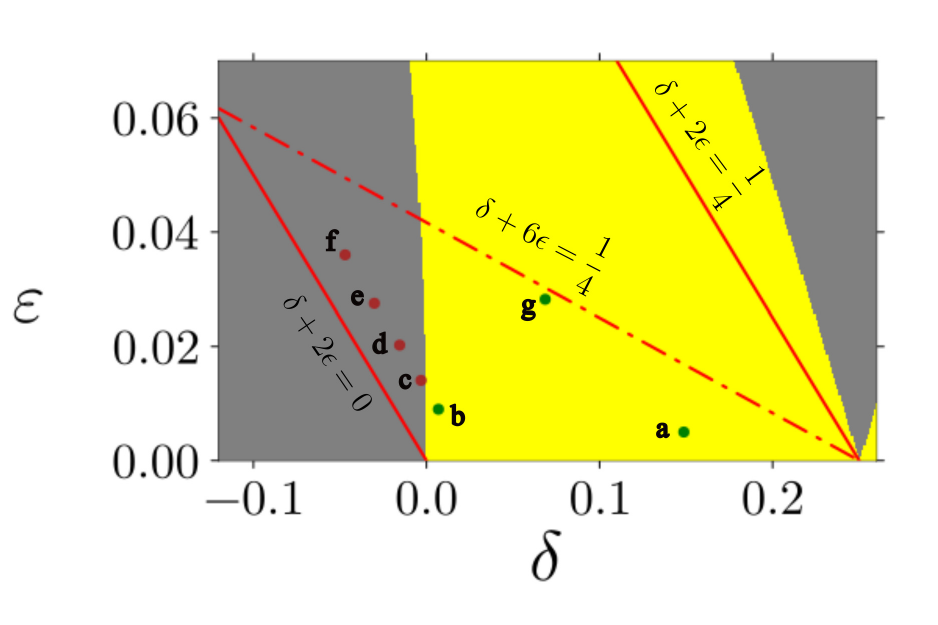}
		\caption{}
	\end{subfigure}
	\captionsetup{width=\columnwidth}
	\caption{(a) The stability chart of the Mathieu equation (\ref{eq9}) denoting tongues of unstable region through Floquet analysis described in appendix B. Grey - Unstable, \textcolor{black}{Yellow} - Stable. Panel (b) Magnified version of the region shown with a rectangle in panel (a). The solid lines, dash-dotted line and points \textbf{a}, \textbf{b}, \textbf{c}, \textbf{d}, \textbf{e} and \textbf{f} have the same meaning as in  fig. \ref{fig:hill_stab}.}. 
	\label{fig:mathieu_tongue}
\end{figure}
In the process, we eliminate the multi-frequency excitation of the Hill equation (\ref{eq8}) leading to the Mathieu equation (\ref{eq9}), which has only two frequencies (viz. the $0^{\text{th}}$ frequency (constant) and the primary frequency $2\omega_0$). 
\begin{align}
	\bm{\ddot}{v} + \dfrac{2k}{m} \left[ 1 - \left(\dfrac{L}{a}\right) - \left(\dfrac{A^2 L}{2a^3}\right) - \left(\dfrac{A^2 L}{2 a^3}\right) \cos(2\omega_0 t) \right]v = 0 \label{eq9}
\end{align}
Let 
\begin{align}
	\delta &\equiv \dfrac{1}{4}\left[1-\dfrac{L}{a}- \dfrac{1}{2}\left(\dfrac{A}{a}\right)^2\left(\dfrac{L}{a}\right)\right], \quad
	\epsilon \equiv \dfrac{1}{16}\left(\dfrac{A}{a}\right)^2\left(\dfrac{L}{a}\right), \quad
	\text{so that } 1 - 4\delta - 8\epsilon = \dfrac{L}{a}.
	\label{eq10}
\end{align}
Note that the overdots in eqns. (\ref{eq9}) represent derivative with respect to $t$ in contrast to eqn. (\ref{eq11}) below where it is with respect to $\tau$ (non-dimensional time). The Mathieu equation (\ref{eq9}) may be compactly rewritten in the standard format using $\epsilon$ and $\delta$ as:
\begin{align}\label{eq11}
	\dfrac{d^2v}{d\tau^2} + \big[\delta - 2\epsilon\cos(\tau)\big]v(\tau) = 0
\end{align}
with $\tau \equiv 2\omega_0 t$.

The stability chart (see fig \ref{fig:mathieu_tongue}) for the Mathieu equation (\ref{eq11}) is obtained similar to the Hill equation (see Appendix  \ref{Floquet}). The region in grey in fig. \ref{fig:mathieu_tongue} is unstable, while that in yellow is stable - the well-known stability tongues of the Mathieu equation are apparent \cite{bender2013advanced}. \textcolor{black}{The same physical constraints on $(L/a)$ and $(A/a)$ as discussed for the Hill equation apply here, restricting the accessible $\delta$-$\epsilon$ values to lie between the two solid-red straight lines in fig. \ref{fig:mathieu_tongue}(a). This region is magnified in fig. \ref{fig:mathieu_tongue}(b); the dash-dotted line $\delta + 6\epsilon=1/4$ (same as in fig. \ref{fig:hill_stab}) corresponds to $|A/a| = 1$, and the space between this and $\delta+2\epsilon=1/4$ is again not permitted.}

It is clear from a comparison of figures \ref{fig:hill_stab} and \ref{fig:mathieu_tongue}b that the Hill equation admits a larger unstable region within the lines $\delta +2\epsilon =0$ and $\delta + 6\epsilon =\dfrac{1}{4}$, compared to the Mathieu equation. We have done a consistency check by solving the Hill differential equation and the Mathieu equation \textcolor{black}{for point (g)} $\delta = 0.0686, \epsilon =0.0282$. This choice of parameters corresponds to a point where the solution to the Mathieu equation is stable while that of the Hill equation is unstable - our numerical solution to both equations validates this prediction for this choice of parameters, we do not provide this data here. Further, we compare the prediction from the Hill and the Mathieu equations with the numerical solution to the full nonlinear eqns. (\ref{eq3}) and (\ref{eq4}) in the next sub-section.
\textcolor{black}{In order to validate these linearised predictions}, we have obtained the solution to the full eqns. (\ref{eq3}) and (\ref{eq4}) numerically, employing the stability charts in figs. \ref{fig:hill_stab} and \ref{fig:mathieu_tongue} to inform us about nature of the solution viz. linearly stable or unstable. \textcolor{black}{This is described next}.
\subsubsection{\textcolor{black}{Results of the full nonlinear equations and its comparison with Hill and Mathieu equations:}}\label{sec:nonlinearevolution} 
\begin{figure}[h]
	\centering
	\begin{subfigure}{0.45\textwidth}
		\centering
		\hspace{-0.7cm}\includegraphics[scale=0.135,trim={2.5 0 0 2cm},clip]{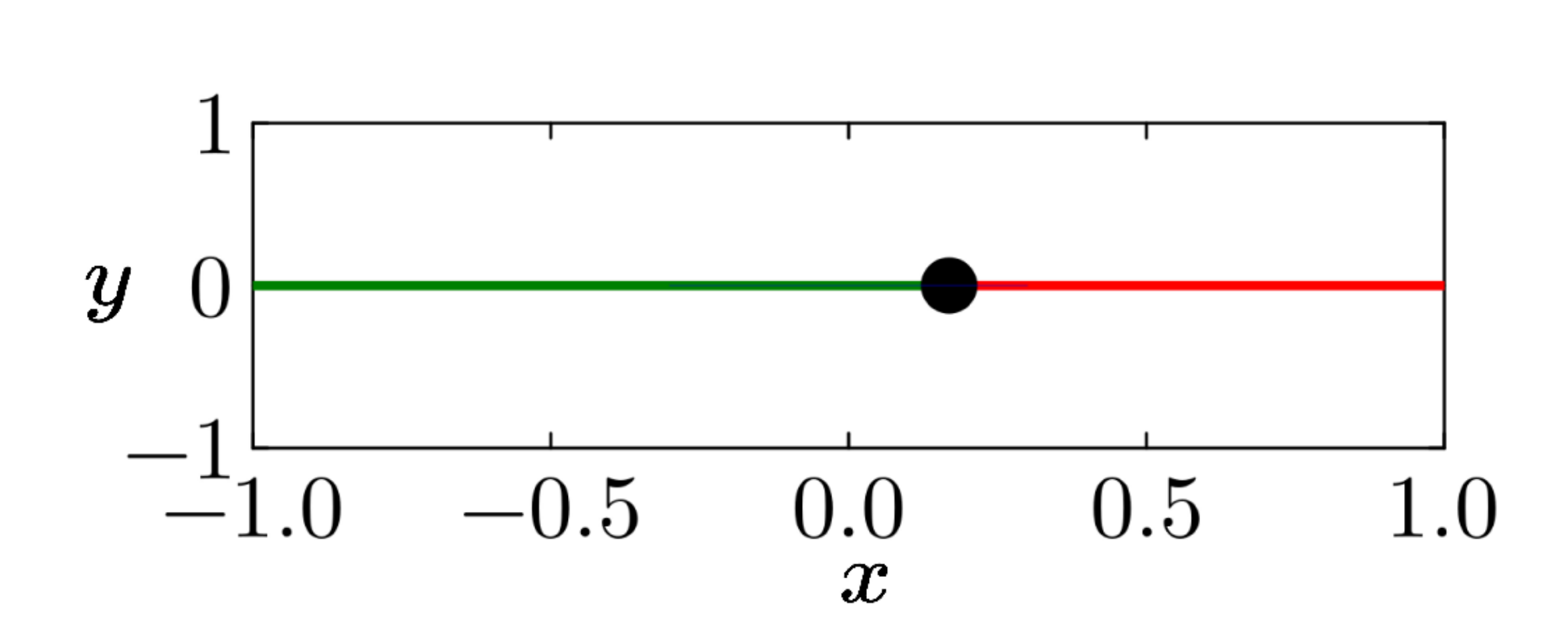}
		\caption{$A = 0.3\;(\delta = 0.1487,\; \epsilon = 0.005)$ - stable (S)}
	\end{subfigure}
	\hfill
	\begin{subfigure}{0.45\textwidth}
		\centering
		\makebox[\textwidth][c]{\includegraphics[scale=0.1,trim={5cm 0 0 2cm},clip]{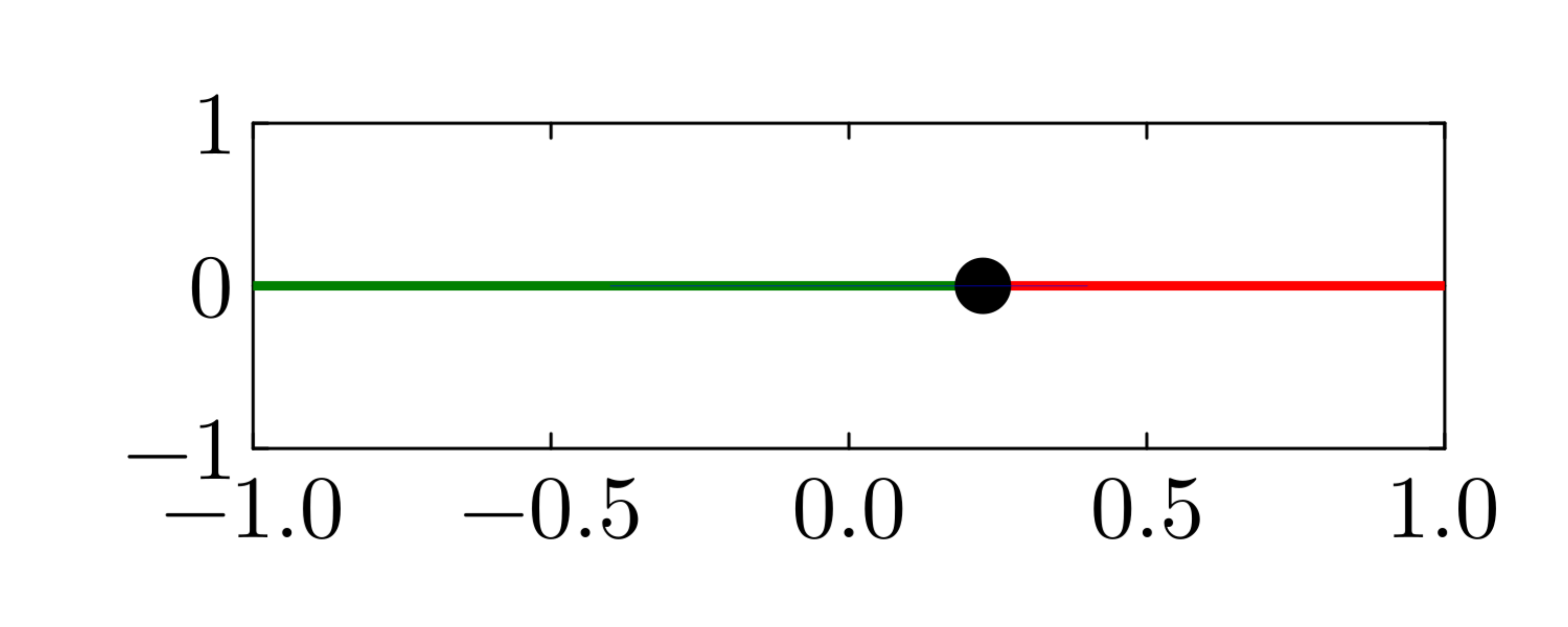}}
		\caption{$A = 0.4\;(\delta = 0.0069,\; \epsilon = 0.009)$ - stable (S)}
	\end{subfigure}
	
	\begin{subfigure}{0.45\textwidth}
		\centering
		\makebox[\textwidth][c]{\includegraphics[scale=0.1,trim={5cm 0 0 2cm},clip]{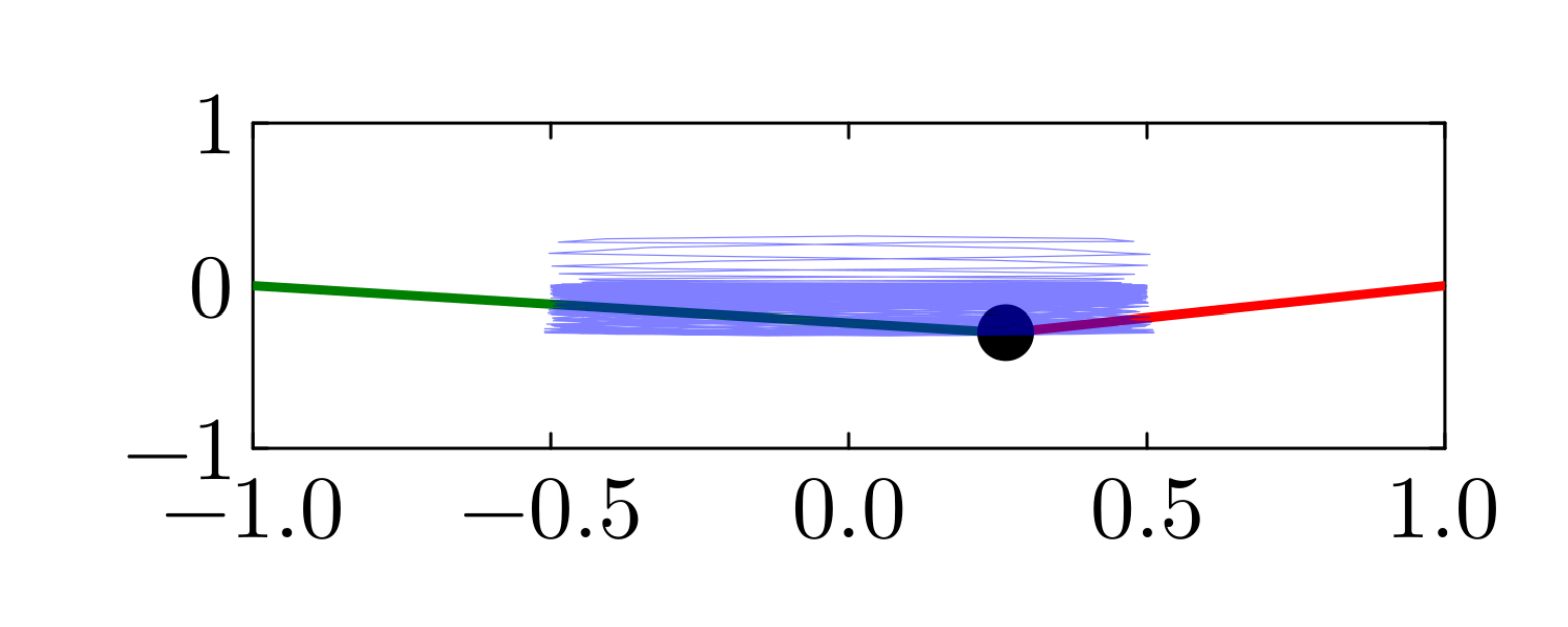}}
		\caption{$A = 0.5\;(\delta = -0.003125,\; \epsilon = 0.01406)$ - unstable (U)}
	\end{subfigure}
	\hfill
	\begin{subfigure}{0.45\textwidth}
		\centering
		\makebox[\textwidth][c]{\includegraphics[scale=0.1,trim={5cm 0 0 2cm},clip]{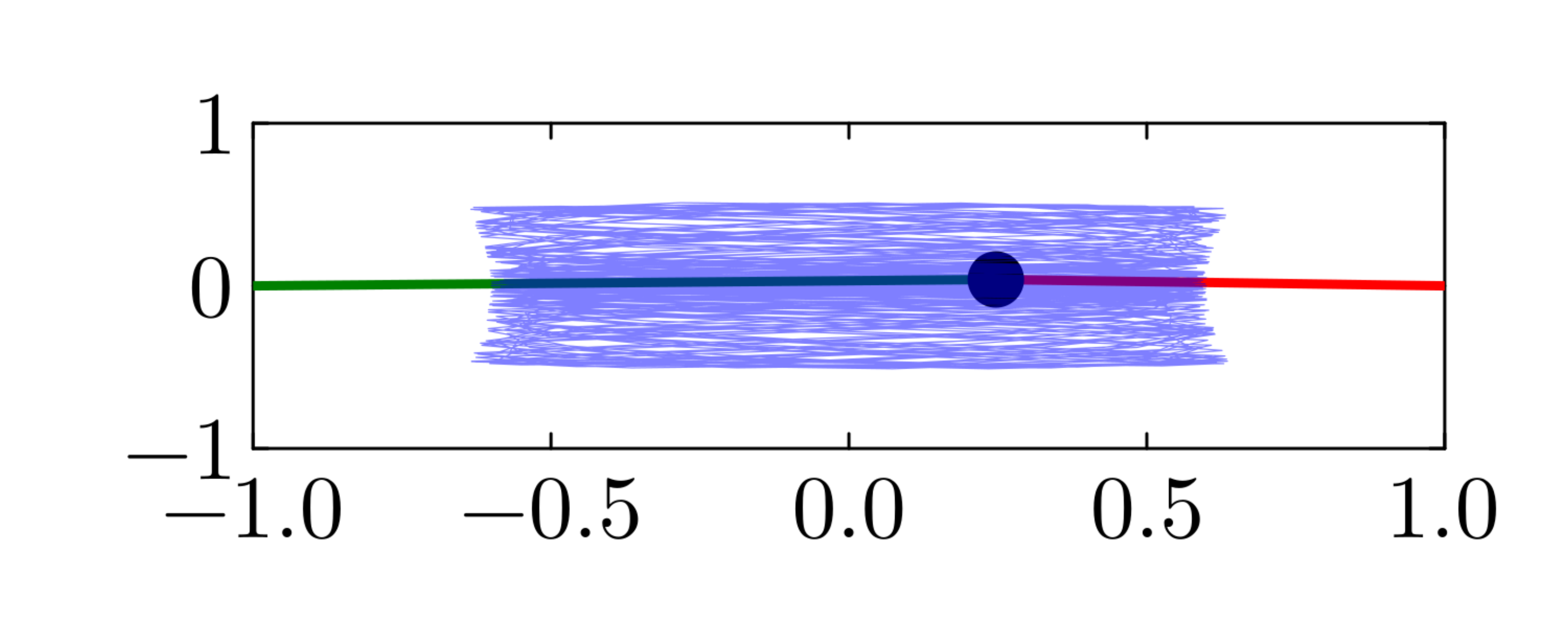}}
		\caption{$A = 0.6\;(\delta = -0.0155,\; \epsilon = 0.02025)$ - unstable (U)}
	\end{subfigure}
	
	\begin{subfigure}{0.45\textwidth}
		\centering
		\makebox[\textwidth][c]{\includegraphics[scale=0.1,trim={5cm 0 0 2cm},clip]{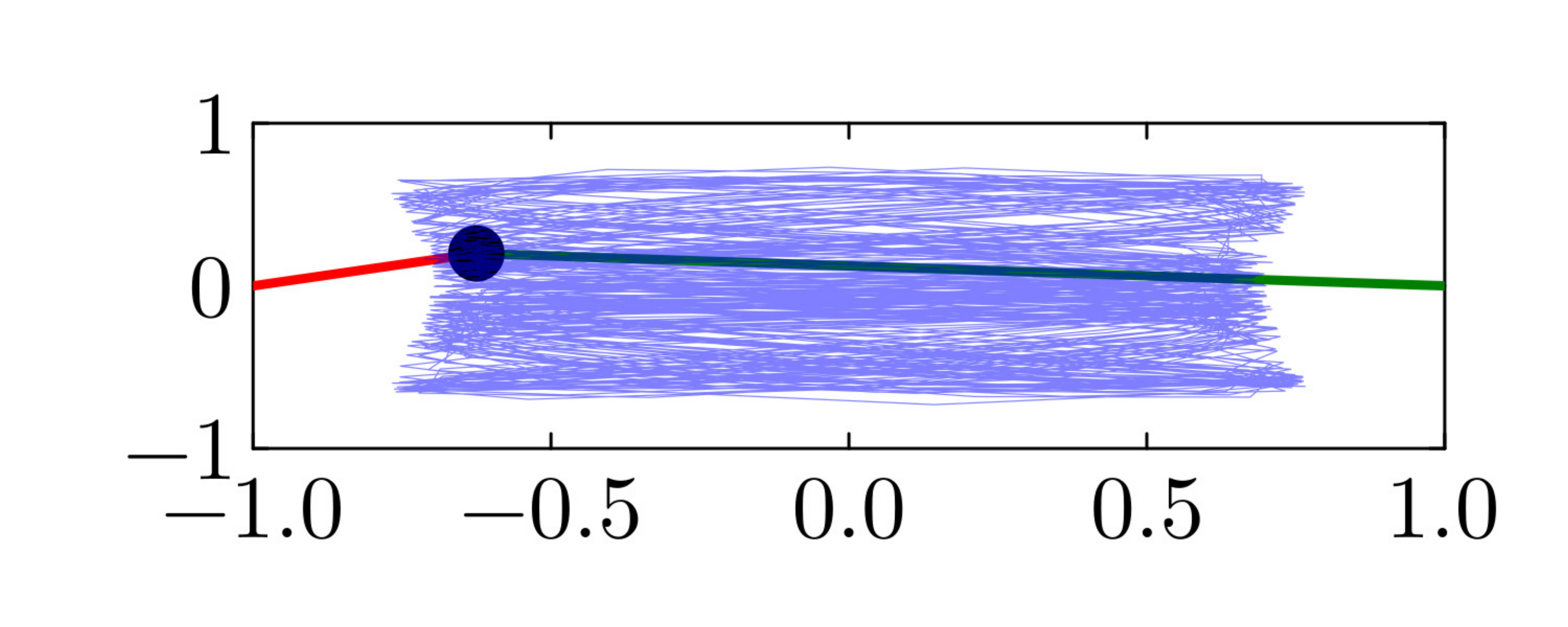}}
		\caption{$A = 0.7\;(\delta = -0.03012,\; \epsilon = 0.0275)$ - unstable (U)}
	\end{subfigure}
	\hfill
	\begin{subfigure}{0.45\textwidth}
		\centering
		\makebox[\textwidth][c]{\includegraphics[scale=0.1,trim={5cm 0 0 2cm},clip]{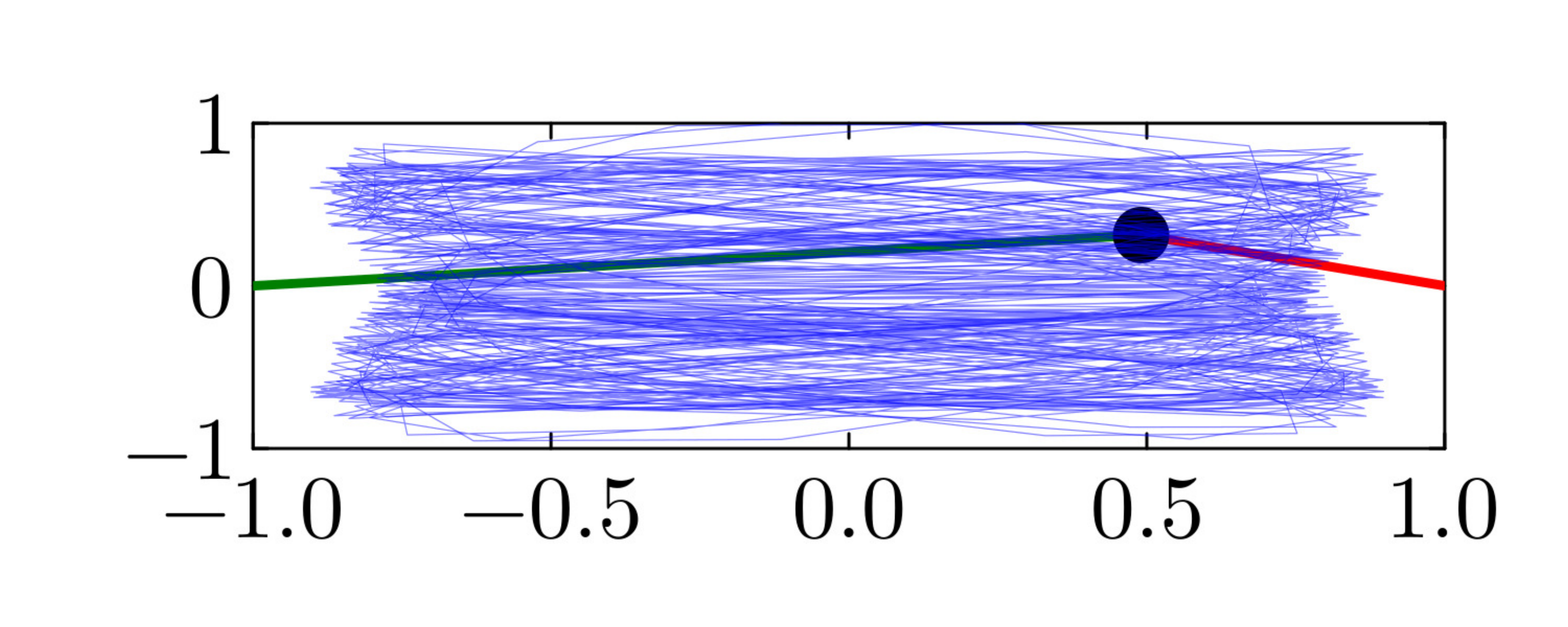}}
		\caption{$A = 0.8\;(\delta = -0.0470,\; \epsilon = 0.036)$ - unstable (U)}
	\end{subfigure}
	\captionsetup{width=\textwidth}
	\caption{Long-time $\left(\dfrac{t\omega_0}{2\pi}\approx 160\right)$ solution of eqns.~(\ref{eq3}) and (\ref{eq4}). The trajectory of the mass $m$ is traced in blue with time on the $x$-$y$ plane for different $x(0)$: (a) $A=0.3$ (S), (b) $A=0.4$ (S), (c) $A=0.5$ (U), (d) $A=0.6$ (U), (e) $A=0.7$ (U), (f) $A=0.8$ (U). The red/green colour denotes spring stress, viz., red for compression and green for tension. Parameters: $L=0.9$, $a=1$, $\omega_0=1$ for all plots. See fig.~\ref{fig:hill_stab} for stability points.}
	\label{fig:numericalSol_Timetrace}
\end{figure}

In figures \ref{fig:numericalSol_Timetrace}(a)-(f), we plot the trajectories of the mass $m$ (indicated as a black circle) traced on the $x-y$ plane with time, obtained by numerically solving eqns. (\ref{eq3}) and (\ref{eq4}).
 This is carried out using DP5 (Dormand-Prince Runge Kutta algorithm) solver from DifferentialEquations.jl \cite{rackauckas2017differentialequations}, an open-source package for solving differential equations in Julia. 
The spring-mass system is initialised with the amplitude $x(0)=A$ and a small perturbation in the vertical coordinate $y(0)=10^{-4}$. In the stable regime i.e. subfigures (a) and (b) in fig. \ref{fig:numericalSol_Timetrace}, this results in pure oscillatory motion in the $x$ direction with no perceptible displacement in the vertical direction, even at long integration time of $\dfrac{t\omega_0
}{2\pi}\approx 160$. 
This is expected behaviour as the parameters for these (cases (a) and (b)) correspond to the stable regime indicated in yellow in \textcolor{black}{figs. \ref{fig:hill_stab} and \ref{fig:mathieu_tongue}}. In contrast, cases (c), (d), (e) and (f) correspond to going progressively deeper into the unstable regime \textcolor{black}{in these stability charts}. As is evident in fig. \ref{fig:numericalSol_Timetrace} subpanels (c)-(e), the numerical solution shows an increasing vertical excursion, reflecting this instability. The significantly larger vertical excursion in case (f) compared to others in the same time window, reflects the larger growth rate for this case. Notice that this case (f) corresponds to the farthest point inside the unstable grey region \textcolor{black}{therein}.

\begin{figure}
	\centering
	\makebox[\textwidth][c]{\includegraphics[scale=0.32,trim={0 0 0 0},clip]{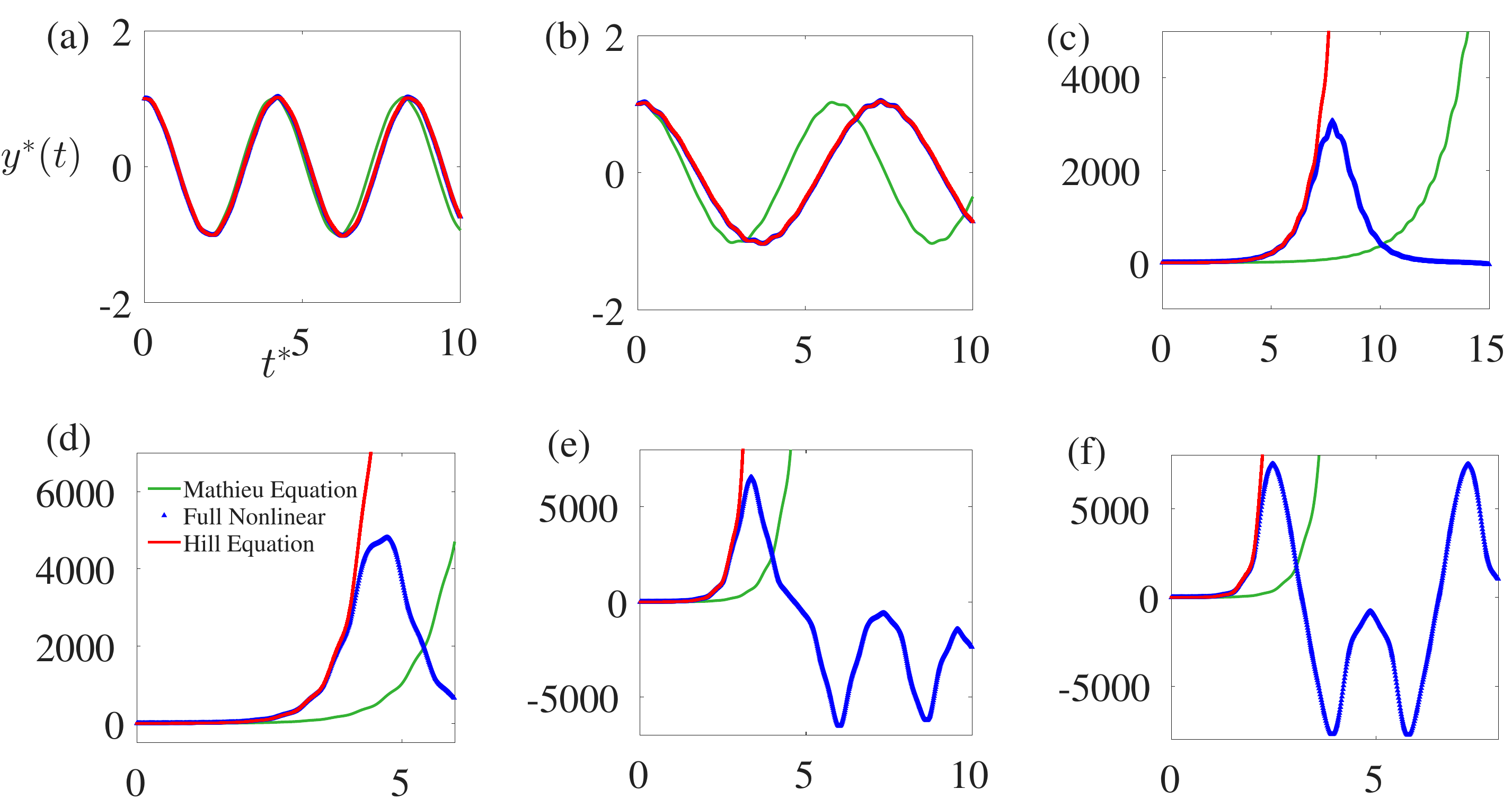}}
	
%
	\captionsetup{width=\textwidth}
	\caption{Vertical displacement $y^{*}(t)\equiv \dfrac{y(t)}{y(0)}$ versus $t^{*} \equiv \dfrac{t\omega_0}{2\pi}$ obtained from numerical solution to equations (\ref{eq3}) and (\ref{eq4}) with the numerical solution to the Hill and the Mathieu equation, eqns. (\ref{eq8}) and (\ref{eq9}) respectively. Both Hill and Mathieu solutions are numerically initialised with $v(0) = 10^{-4}$, the same value used for the full numerical equations. \textcolor{black}{Panels (a)-(f) are parametrized in same manner as corresponding panels in fig. \ref{fig:numericalSol_Timetrace}.}\label{fig:comp}}
\end{figure}
To facilitate further comparisons with the Hill equation solution (eq. (\ref{eq8}) ) and the Mathieu solution (eq. \ref{eq9}), we also compare the vertical displacement of the mass $y^{*}(t)\equiv \dfrac{y(t)}{y(0)}$ as a function of time $t^{*} \equiv \dfrac{t\omega_0}{2\pi}$. This is depicted in fig. \ref{fig:comp}. The parameters for each of these subpanels are the same as the corresponding subpanels in fig. \ref{fig:numericalSol_Timetrace}. The first two panels correspond to the stable regime. Interestingly, the Mathieu equation solution (green) starts deviating from the full non-linear solution (blue dots) rather early in time. In contrast, the solution to the Hill equation (red), which accounts for higher frequencies in the Fourier series, agrees far better than the Mathieu solution. This behaviour is further enhanced in the unstable regime where exponential growth is expected. This is depicted in panels (c)-(f) in fig. \ref{fig:comp}. Importantly, both the Hill and the Mathieu solution diverge exponentially. This unphysical behaviour at large time is rectified by nonlinear contributions in the full nonlinear solution (blue dots) which display non-linear saturation following a brief period of exponential growth.

\textcolor{black}{To summarize, the time-periodic solution of the mechanical oscillator ($x_b(t)=A\cos(\omega_0 t),\, y_b(t)=0$) is linearly stable for small $A$ but becomes unstable to vertical perturbations beyond a critical amplitude, as predicted by the Hill and Mathieu stability charts (figs. \ref{fig:hill_stab} and \ref{fig:mathieu_tongue}). The Hill equation captures the short-time dynamics accurately (fig. \ref{fig:comp}), beyond which nonlinear effects dominate. With this background, we now return to the interfacial oscillator to establish the proposed analogy.}
\subsection{The analogy - trivial and time-periodic solutions for the interfacial oscillator and their stability}\label{sec:interfacialoscillator}
We now formally establish the analogy between the mechanical and interfacial oscillator. For this, we examine trivial and time-periodic solutions to eqns. (\ref{eq12})-(\ref{eq16}) and examine their stability.
\subsubsection{Trivial solution for the interfacial oscillator:}\label{sec:trivial}
It is an easy exercise to check that eqns. (\ref{eq12})-(\ref{eq16}) admit the trivial solution $y_s(x,t)=0$ and $\Phi(x,y,t)=0$ (indicating water is quiescent with a flat interface). This is the analogue of the trivial solution $x_b(t)=y_b(t)=0$ in our mechanical oscillator example, earlier. The linear stability of this trivial state is carried out using the expansion $\Phi(x,y,t) = 0 + \phi(x,y,t)$ while the deformed interface is represented by $y_s(x,t) = 0 + \eta(x,t)$. Linearising about the perturbation variables $\phi(x,y,t)$ and $\eta(x,t)$ (employing Taylor series expansions about $y=0$) in eqns. (\ref{eq12})-(\ref{eq16}), we find the following equations governing the perturbation variables $\phi$ and $\eta$:
\begin{align}
	& \nabla^2\phi = 0, \; \left(\dfrac{\partial \phi}{\partial x}\right)_{x = \pm \dfrac{\pi}{k}} = 0, \; \left(\dfrac{\partial \phi}{\partial y}\right)_{y \to -\infty} \to 0, \label{eq17}\\
	& \dfrac{\partial\eta}{\partial t}  - \left(\dfrac{\partial\phi}{\partial y}\right)_{y=0} = 0, \label{eq18}\\
	& \left(\dfrac{\partial\phi}{\partial t}\right)_{y=0} + g\eta = 0. \label{eq19}
\end{align}
One can combine eqns. (\ref{eq18}) and (\ref{eq19}) to obtain a single equation at $y=0$. This is:
\begin{align}
	\left(\dfrac{\partial^2\phi}{\partial t^2}\right)_{y=0} + g\left(\dfrac{\partial\phi}{\partial y}\right)_{y=0} = 0. \label{eq20}
\end{align}
We know that the Laplace equation (\ref{eq17}) admits standing wave solutions of the form $\phi(x,y,t) = \alpha(t ; k)\exp(ky)\cos(kx)$ for any $\alpha(t;k)$ and real $k$. By construction, this choice of solution to the Laplace equation also satisfies the two boundary conditions in eqn. (\ref{eq17}), as can be checked readily. Substituting this form for $\phi(x,y,t)$ into eqn. (\ref{eq20}), we find the harmonic oscillator equation governing $\alpha(t;k)$ viz. 
\begin{align}
	\bm{\ddot}{\alpha} + \omega^2\alpha(t;k) = 0, \label{eq21}
\end{align}
where $\omega^2\equiv gk$. This solution represents a standing wave at the interface, which evolves as:
\begin{align}
	\eta(x,t) = a_1\cos(kx)\sin(\omega t), \label{eq22}
\end{align}
where $a_1$ is related to the constant of integration in the solution to eqn. (\ref{eq21}). Our exercise so far parallels the linear stability analysis of the trivial solution obtained earlier; This may be seen by comparing eqn. (\ref{eq21}) with eqns. (\ref{eq5}) and (\ref{eq6}). In the case of the interfacial oscillator on a horizontally confined domain, $k$ is expected to take values from a countably infinite set of numbers and thus eqn. (\ref{eq21}) is the equation governing the temporal evolution of each such admissible value of $k$ - this simply reflects the fact that our interfacial oscillator has infinite degrees of freedom. In contrast, eqns. (\ref{eq5}) and (\ref{eq6}) reflect the two degrees of freedom (\textcolor{black}{x, y}) of the mechanical oscillator.
\subsubsection{Time-periodic solution for the interfacial oscillator:}\label{sec:nontrivial}
Similar to the mechanical oscillator equations, one can analogously enquire if there are time-periodic solutions to the full set of partial differential eqns. (\ref{eq12})-(\ref{eq16}). However, this is a point of departure for this analogy, as far as mathematical difficulty is concerned. Spotting an exact solution to the mechanical oscillator eqns. (\ref{eq3}) and (\ref{eq4}) was quite easy \textcolor{black}{and may even be anticipated physically}. However, the interfacial oscillator is governed by a far more complicated set of partial differential eqns. (\ref{eq12})-(\ref{eq16}) along with nonlinear boundary conditions. It so turns out that these equations too admit \textcolor{black}{time-periodic solutions \cite{strutt1915deep} which can be linearly unstable \cite{Mercer1992}; however, it requires extensive calculations to even find these solutions. Improving on Rayleigh's analysis \cite{strutt1915deep}, \citet{Penney1952} obtained a fifth order ($\mathcal{O}(\hat{A}^5)$, $\hat{A}\equiv a_1k$) representation of such a solution to eqns. (\ref{eq12})-(\ref{eq16}); for further developments, see \cite{schwartz1981semi,saffman1979note,rottman1982steep}. To compare with their solution,} we re-write expression (\ref{eq22}) as :
\begin{align}
	\tilde{\eta} \equiv k\eta = \hat{A}\cos(kx)\sin(\omega t), \quad \hat{A}\equiv ka_1 \label{eq23}
\end{align}%

The expression \textcolor{black}{of \citet{Penney1952}} for the shape of a standing-wave $\tilde{\eta}$ of finite steepness $\hat{A}$ is given by eqn (\ref{eq2}).  
\begin{figure}
	\centering
 \subfloat[][
]{
	\includegraphics[scale=0.22,trim={0 0 0 0cm},clip]{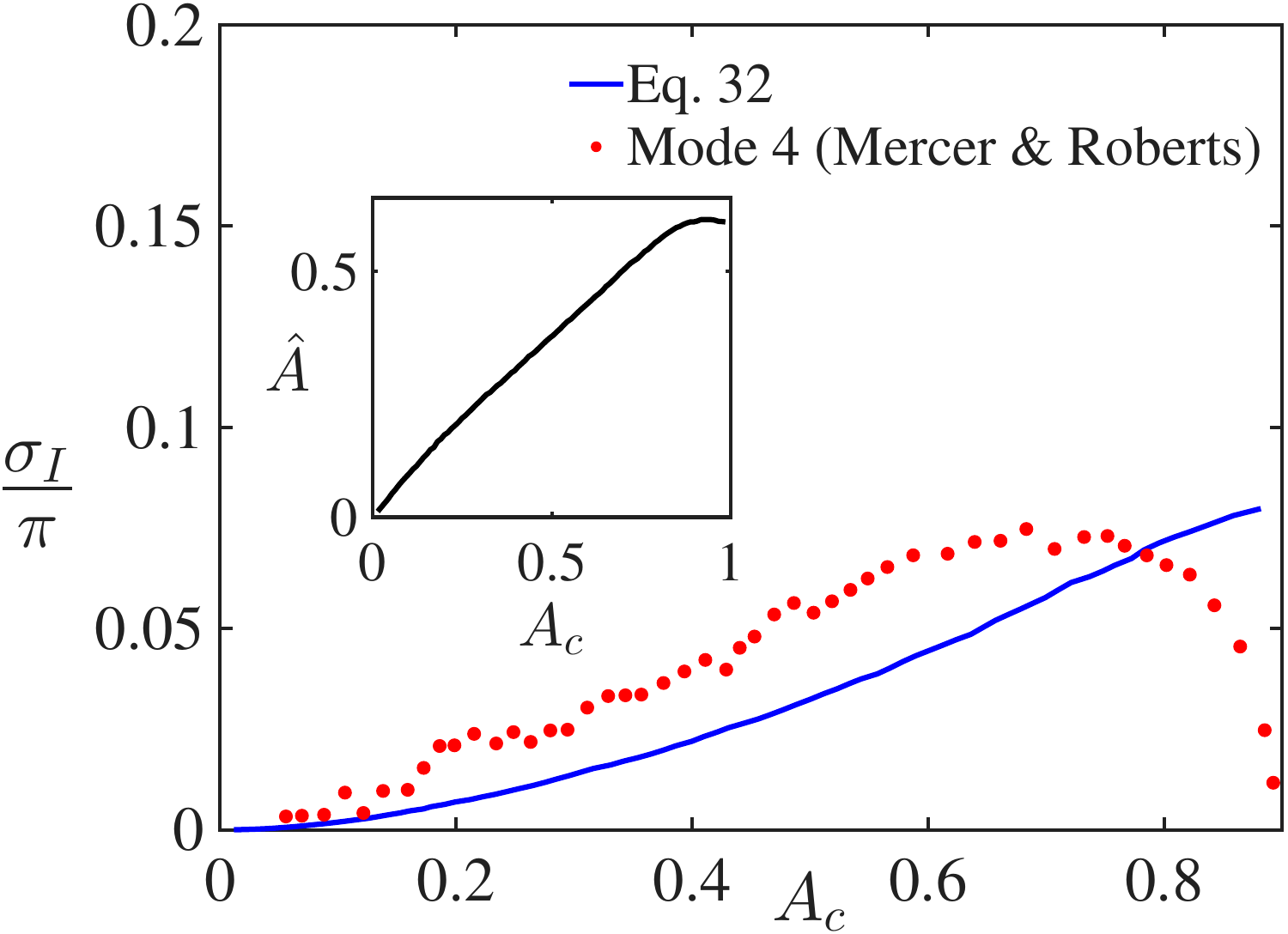}
	\label{fig:mercer_roberts}
}\hspace{1cm}
\subfloat[][
]{
	\includegraphics[scale=0.22,trim={0 0 0 0cm},clip]{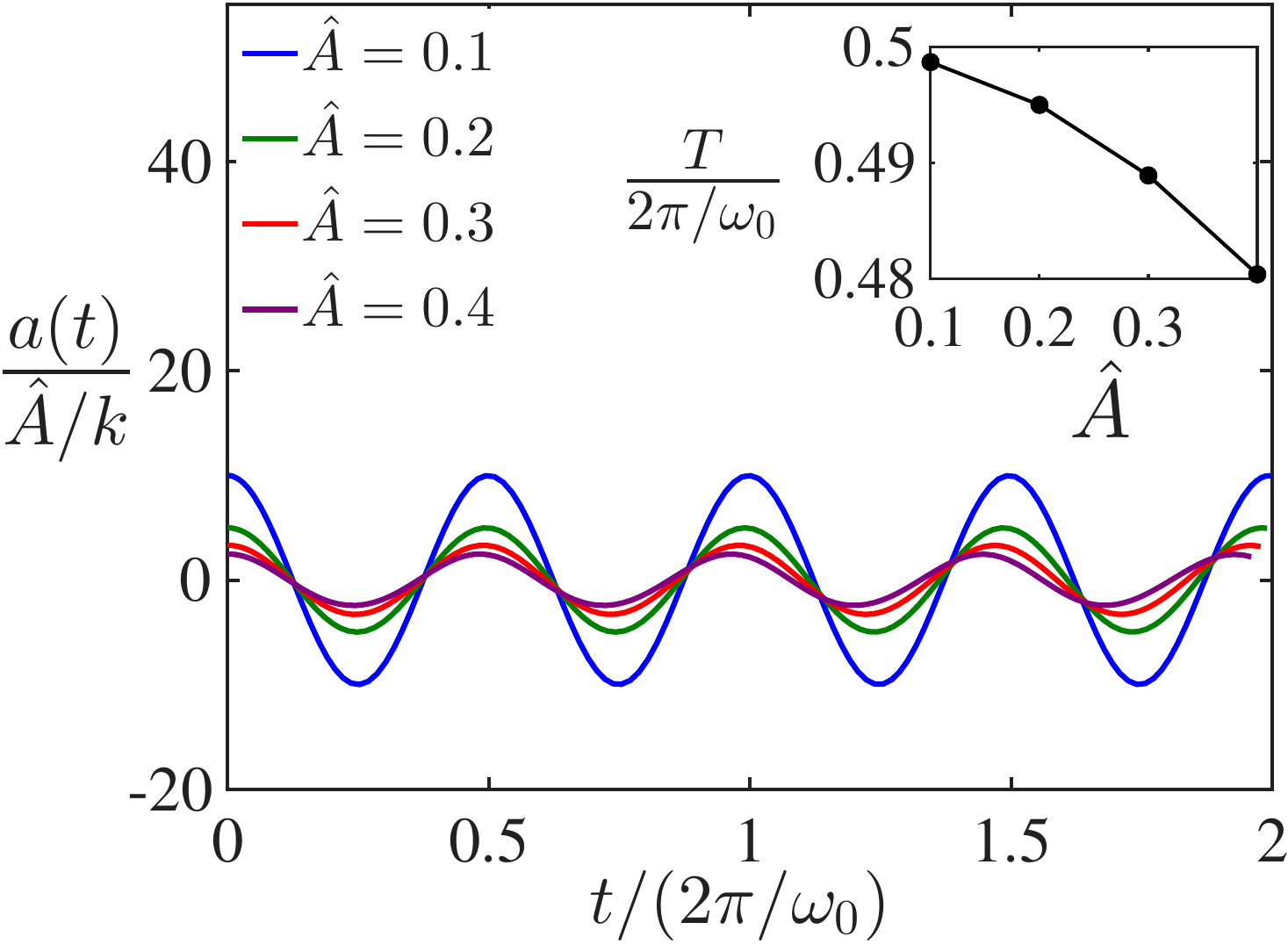}
	\label{fig:mathieulike}
}
	\captionsetup{width=\textwidth}
	\caption{ \textcolor{black}{Panel (a) Eigen-frequencies ($\sigma_I$) of a superharmonic mode versus crest acceleration ($A_c$). The crest acceleration can be related to the base-state wave steepness $\hat{A}$ as shown in the inset based on data extracted from \cite{Mercer1992}. Red dots denote the eigen-frequencies calculation in \cite{Mercer1992}, while the solid blue line shows a qualitative agreement (in small to moderate $\hat{A}$ range) from reduced-order interfacial oscillator stability model of eqn. \ref{eq33}.} \textcolor{black}{ Panel (b) Numerical solution of eqn. (\ref{eq33}) for different values of $\hat{A}$ for $a(0)=1,\; a'(0)=0$.  Increasing $\hat{A}$ affects the frequency, \textcolor{black}{note the decrease in time-period with increase in $\hat{A}$ in the inset.}}}
\end{figure}
We observe that as $\hat{A}\rightarrow 0$, expression (\ref{eq2}) reduces to that in (\ref{eq23}) at leading order, thereby validating linearized predictions. \textcolor{black}{Like the mechanical oscillator, the solution \ref{eq23} too contains a free parameter ($\hat{A}$). In the case of the interfacial oscillator however, this parameter dictates the shape and frequency of oscillation, and its value determines the (linear) stability of the nonlinear oscillation.} 
This linear stability analysis was first carried out by \citet{Mercer1992} who computed the finite-amplitude, time-periodic, standing waves numerically with high accuracy and then evaluated its stability using Floquet analysis\textcolor{black}{, refer to equation $(21)$ therein}. In the remainder of this section, as well as the next section, we will demonstrate that some of results of Mercer 1992, may be obtained using simplified models, discussed so far. Towards this, it is useful to briefly summarise qualitative aspects of their stability results first.

\textcolor{black}{Floquet analysis from \cite{Mercer1992} yields, \textcolor{black}{ countably infinite set of complex eigenvalues $\sigma = \sigma_R + i \sigma_I $. By construction $\sigma_R$ corresponds to the growth rate and $\sigma_I$ corresponds to frequency of the oscillation, the latter is plotted in fig \ref{fig:mercer_roberts} as red dots obtained from \cite{Mercer1992}}. Eigenmodes are classified as being members of two families viz. subharmonic or super-harmonic. This nomenclature is set by the wavelength of the longest eigenmode in that family. For the super-harmonic family (our focus here), the longest eigenmode is equal to or shorter than the wavelength of the standing wave \cite{LonguetHiggins1978,Mercer1992}.
Note that the individual eigenmodes although periodic in space are not single sine or cosine functions and require computing a Fourier series to represent them.  The label \say{Mode $ 4$} in fig. \ref{fig:mercer_roberts} indicates that wavelength of this mode is one-fourth of the standing wave, when its steepness is sufficiently
low.} It is seen that this mode is stable up to a large steepness of $\hat{A} \approx 0.592$ (see figure \ref{fig:mercer_roberts} inset). \textcolor{black}{In the following section, using ideas discussed so far, we show that a reduced-order simplified model, also leads to similar conclusions.}
 
\subsubsection{Linear stability analysis in \textcolor{black}{an over-simplified model}}\label{sec:stabilityInterfacial}
\textcolor{black}{In line with the spirit of the analogy, we derive here a novel Mathieu-like equation for the stability of the interfacial oscillator using a highly truncated model, both for the standing wave (base-state) and the eigenmode (perturbation). Our aim is to obtain equations similar to the Mathieu equation derived earlier, for the spring-mass system.} As we seek to highlight the analogy between the two systems, we do not strive for very accurate linear stability results. Consequently, in what follows we represent the base-state (the nonlinear standing wave whose stability is sought), employing a two-term (dimensional) representation upto $\mathcal{O}(\hat{A}^2)$ \textcolor{black}{\cite{Penney1952}} :
	\begin{align}
		&\eta_b(x,t) = \dfrac{\hat{A}}{k}\sin(\omega_0 t)\cos(kx) + \dfrac{1}{4k}\hat{A}^2 \Biggl(1 - \cos(2\omega_0 t) \Biggr)\cos(2kx)+\mathcal{O}(\hat{A}^3),  \label{eq25}\\
		& \phi_b(x,y,t) = \dfrac{\hat{A}}{k}\sqrt{\dfrac{g}{k}}\cos(kx)e^{ky}\cos(\omega_0 t) + \mathcal{O}(\hat{A}^3), \label{eq26}
	\end{align}
	where $\omega_0 = \sqrt{gk}\sqrt{(1 - \hat{A}^2/4)}$ and $\hat{A}$ is a non-dimensional amplitude, equivalent to $a_1k$ in eqn. (\ref{eq23}). The understanding is that the infinite series version of eqn. (\ref{eq25}) and (\ref{eq26}) satisfy equations (\ref{eq15})-(\ref{eq16}) exactly, by definition. 

\textcolor{black}{Perturbing this base-state using the form (\ref{eqc1})-(\ref{eqc2}) in Appendix (\ref{app3}), the perturbations $\tilde{\eta}, \tilde{\phi}$ are also chosen to have the following truncated (dimensional) form:
	\begin{align}
		&\tilde{\eta}(x,t) = a(t)\cos(4kx) + \ldots\label{eq27}\\
	&	\tilde{\phi}(x,y,t) = b(t)\cos(4kx) e^{4ky} + \ldots \label{eq28}
	\end{align}
	Two comments are necessary: firstly, the $x$ dependence of the first terms in (\ref{eq27}) \& (\ref{eq28}) represent the leading order behaviour in a Fourier-series representation of the spatial form of the perturbation. We retain only the first terms in these series, so as to simplify subsequent algebra with an intuitive expectation that such a low-order representation (of both base-state and perturbation) might still yield \textcolor{black}{to a} useful model.}

\textcolor{black}{Secondly, the choice of $4kx$ (perturbation wavelength one-fourth of the base-state), is due to the fact that fig. \ref{fig:mercer_roberts} shows that the frequency of the fourth mode can coalesce with that of the primary mode (base-state wavelength) at a relatively large value of $\hat{A}$ to generate super-harmonic instability, of interest to us here. Due to the truncated nature of both base and perturbation representations, it will be seen that the resultant equation for $a(t)$ in eqn. (\ref{eq27}) will predict stable behaviour, consistent with fig. \ref{fig:mercer_roberts} for $\hat{A}<0.4$, the regime where our truncated model can be expected to be a reasonable approximation.} 

\textcolor{black}{Plugging equations (\ref{eq25})-(\ref{eq28}) into equation (\ref{eqc6}) obtained from the kinematic boundary condition appendix (\ref{app3}), we obtain :
	\begin{align}
		& a'(t)\cos(4kx) + a(t)\omega\hat{A}\cos(\omega_0 t)\Biggl[\dfrac{3}{2}\cos(3kx) -  \dfrac{5}{2}\cos(5kx) + \dfrac{\hat{A}}{2}\cos(2kx)\sin(\omega_0 t) - \nonumber \\
		&\quad \dfrac{3\hat{A}}{2}\cos(6kx)\sin(\omega_0 t)\Biggr] - 4 k b(t)\cos(4kx) -  k b(t) \sin(\omega_0 t)\Biggl[6\hat{A}\cos(3kx) + 10\hat{A}\cos(5kx) + \nonumber \\
		&\quad 6\hat{A}^2\cos(2kx)\sin(\omega_0  t) + 16\hat{A}^2 \cos(4kx)\sin(\omega_0 t) + 18\hat{A}^2\cos(6kx)\sin(\omega_0 t)\Biggr]= 0. \label{eq29}
	\end{align}
	Similarly, plugging equations (\ref{eq25})-(\ref{eq28}) into equation (\ref{eqc8}) obtained from the Bernoulli equation in appendix (\ref{app3}), we obtain ($a'(t) \equiv \dfrac{da}{dt}$ and so on) :
	\begin{align}
		& b'(t)\cos(4kx)\Biggl(1 + 4\hat{A}\cos(kx)\sin(\omega_0 t) + 4\hat{A}^2\sin^{2}(\omega_0 t) + 6\hat{A}^2\cos(2kx)\sin^{2}(\omega_0 t)\Biggr) + \nonumber \\
		&\quad b(t)\omega\cos(3kx)\cos(\omega_0 t)\Biggl(4\hat{A} +  20\hat{A}^2 \cos(kx)\sin(\omega_0 t)\Biggr) +  a(t)g \cos(4kx) \Biggl(1 + \nonumber\\
		&\quad  \hat{A}^2 \cos^2(\omega_0 t) -\hat{A}\dfrac{\omega_0}{\omega}\cos(kx)\sin(\omega_0 t)-  \hat{A}^2\dfrac{\omega_0}{\omega}\cos^2(kx)\sin^2(\omega_0 t)\Biggr)= 0. \label{eq30}
	\end{align}
	In order to eliminate the $x$ dependency in (\ref{eq29}) and (\ref{eq30}), we take the inner-product (integrate) of equations (\ref{eq29}) and (\ref{eq30}) with $\cos(4kx)$ over a wavelength $\left(\dfrac{2\pi}{k}\right)$ which leads to the following coupled linear equations for $a(t)$ and $b(t)$  viz :
	\begin{equation}
		\begin{split}
			&a(t)g \Biggl( 1 + \dfrac{\hat{A}^2}{2} \left( 1 - \dfrac{\omega_0}{2\omega} + \left( 1 + \dfrac{\omega_0}{2\omega} \right) \cos(2\omega_0 t) \right) \Biggr) + b'(t) \hat{A}^2 \Biggl( 1 + 4\hat{A}^2 \sin^2(\omega_0 t) \Biggr) \\
			&\quad + 5b(t)\omega \hat{A}^2 \sin(2\omega_0 t) = 0
		\end{split}
		\label{eq31}
	\end{equation}
	and
	\begin{align}
		& a'(t) - 4kb(t)\Biggl(1 + 2\hat{A}^2\ - 2\hat{A}^2 \cos(2\omega_0 t)\Biggr) = 0.  \label{eq32}  
	\end{align}
	Eliminating $b(t)$ and $b'(t)$ (algebra not shown) from the equations  (\ref{eq31}-\ref{eq32}) results in the following Mathieu-like equation upto $O(\hat{A}^2)$:
	\begin{align}
		& a''(t) +  \hat{A}^2 \omega \sin(2\omega_0 t)a'(t) +\omega^2 \Biggl[4 + \hat{A}^2 + 3\hat{A}^2 \cos(2\omega_0 t)  \Biggr]a(t)  = 0. \label{eq33}
	\end{align}
}
\textcolor{black}{\noindent Note the presence of a first derivative term $a'(t)$ with periodic coefficient in eqn. (\ref{eq33}). Instead of carrying out Floquet analysis on (\ref{eq33}), we have chosen to solve it numerically using Runge-Kutta$45$ in Julia \cite{rackauckas2017differentialequations} and results are presented in fig. \ref{fig:mathieulike}.}

Fig \ref{fig:mathieulike} indicates a stable behaviour for super-harmonic perturbation of small-amplitude. The frequency is twice that of the base-state for small $\hat{A}$, as seen in this figure. \textcolor{black}{This is also evident from eqn. \ref{eq33} for the limit $\hat{A} \to 0$, leading to a simple harmonic oscillator $a''(t) + 4\omega^2 a(t) = 0$. For finite $\hat{A}$, the parametric terms involving $\sin(2\omega_0 t)$ and $\cos(2\omega_0 t)$ perturb this frequency, resulting in a small shift $\sigma_I = \omega_{actual} - 2\omega$ which grows with $\hat{A}$. This is corroborated by the qualitative agreement of the temporal frequency (normalised by $\pi$) from analysing eqn. $\ref{eq33}$ and data-extracted for mode $4$ from \cite{Mercer1992}}. The stable behaviour for the range of $\hat{A}$ is expected, as we have employed a two-term representation to the base-state in eqn. (\ref{eq25}) and (\ref{eq26}), expected to be accurate only for moderate values of $\hat{A} \lessapprox 0.4$ (denoted by dotted line in figure \ref{fig:mercer_roberts}). The stable behaviour observed in fig. \ref{fig:mercer_roberts} is consistent with predictions of fig. \ref{fig:mathieulike} in this regime. One also expects that these predictions can be systematically improved by using more accurate representations of base-state as well as perturbation, to recover the instability in fig. \ref{fig:mercer_roberts} at $\hat{A}\approx 0.592$ \textcolor{black}{(corresponding to the maximum amplitude (see fig \ref{fig:standing_waves_0_2T}b))} at the cost of tedious algebra. \textcolor{black}{The simplified analysis here may be extended to include other superharmonic modes, all of which are stable at the steepness indicated in fig. \ref{fig:mercer_roberts}.}	

\begin{figure*}[htbp]
	\centering
	\subfloat[$\dfrac{t}{T}=0$]{\includegraphics[scale=0.27,trim={0.8cm 0 0 0 }]{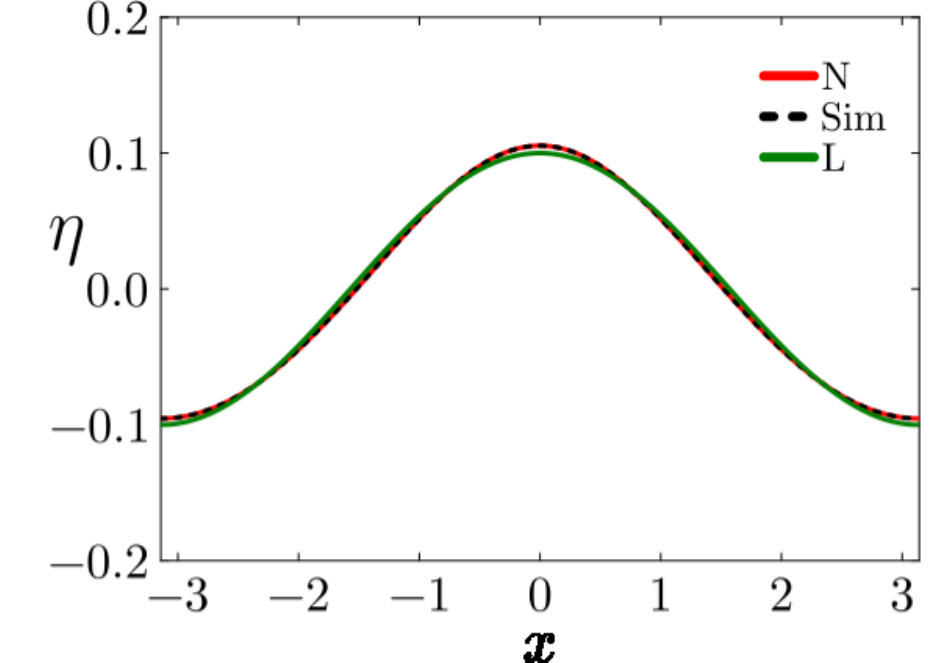}\label{fig:9(a)}}
	\subfloat[$\dfrac{t}{T}=0.25$]{\includegraphics[scale=0.22]{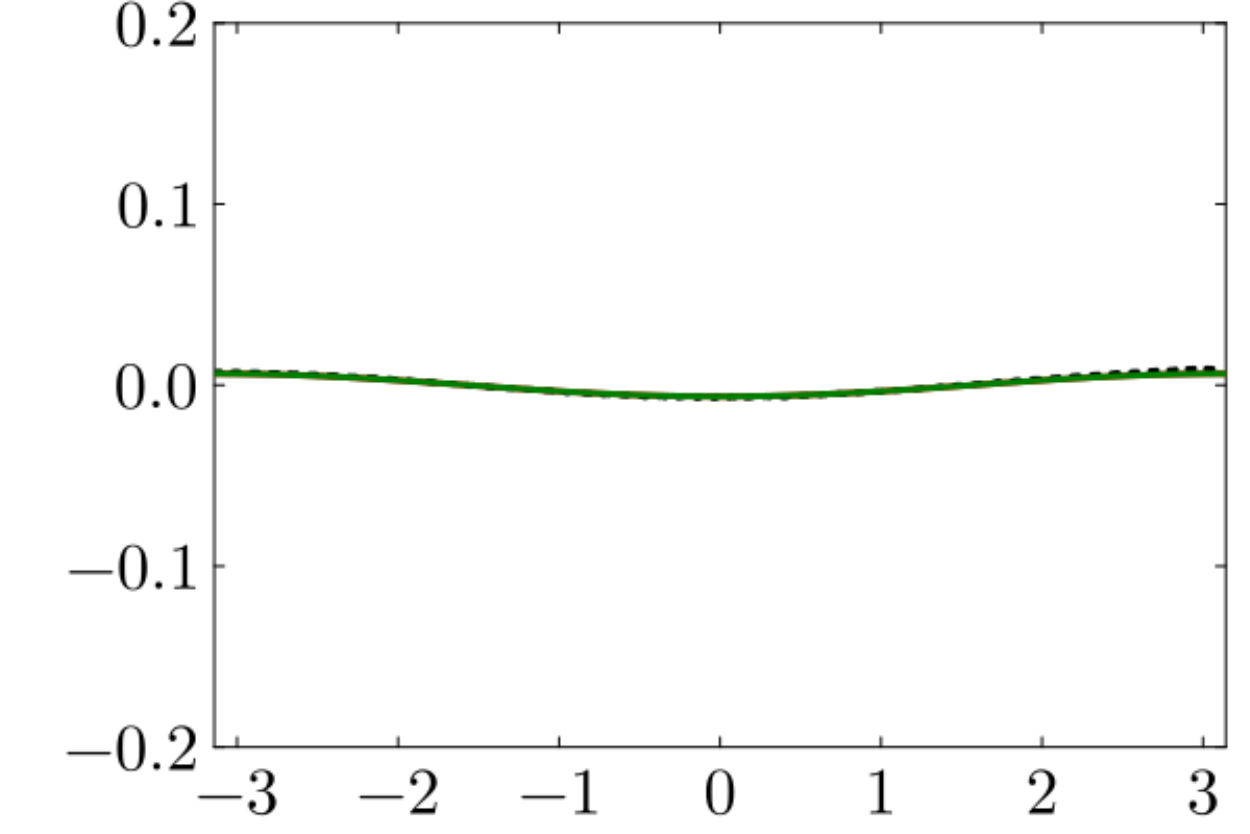}}
	\subfloat[$\dfrac{t}{T}=0.5$]{\includegraphics[scale=0.22]{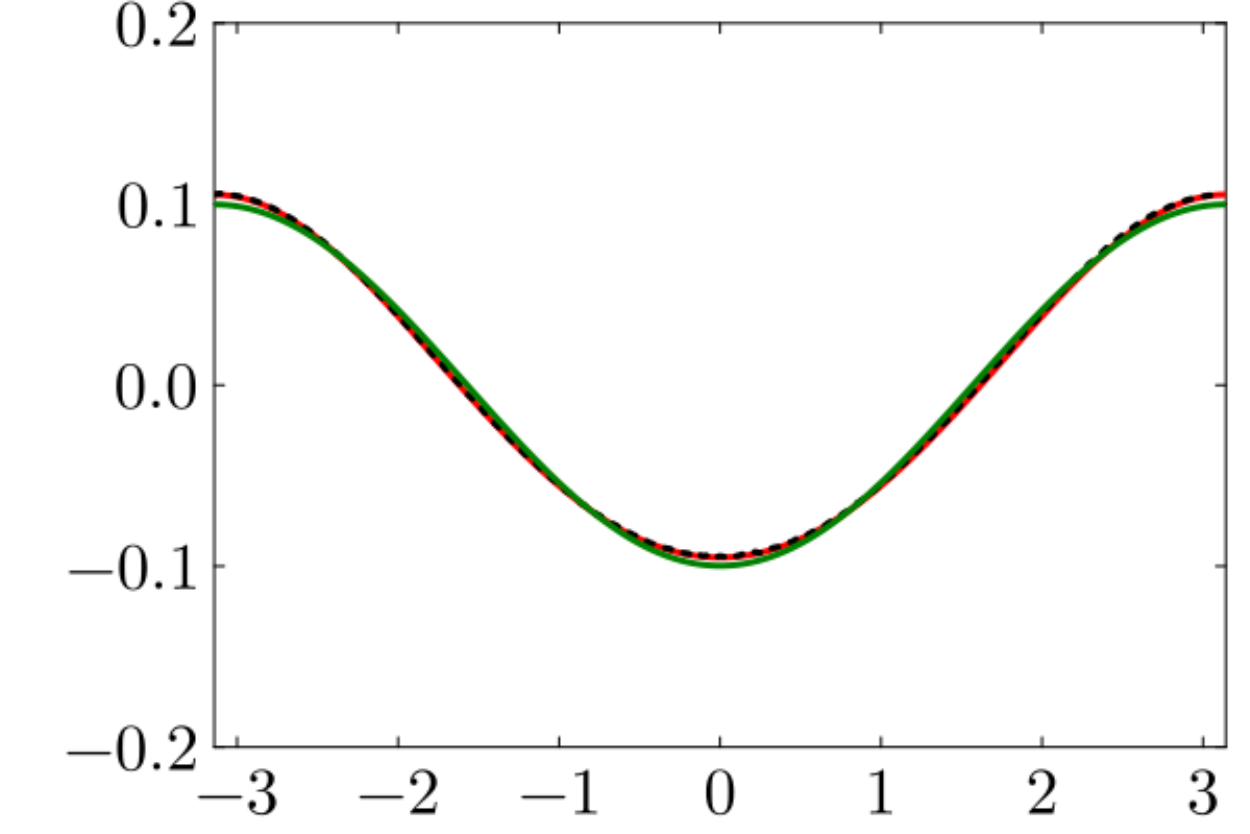}}\\
	\subfloat[$\dfrac{t}{T}=0.75$]{\includegraphics[scale=0.22, trim={0.8cm 0 0 0}]{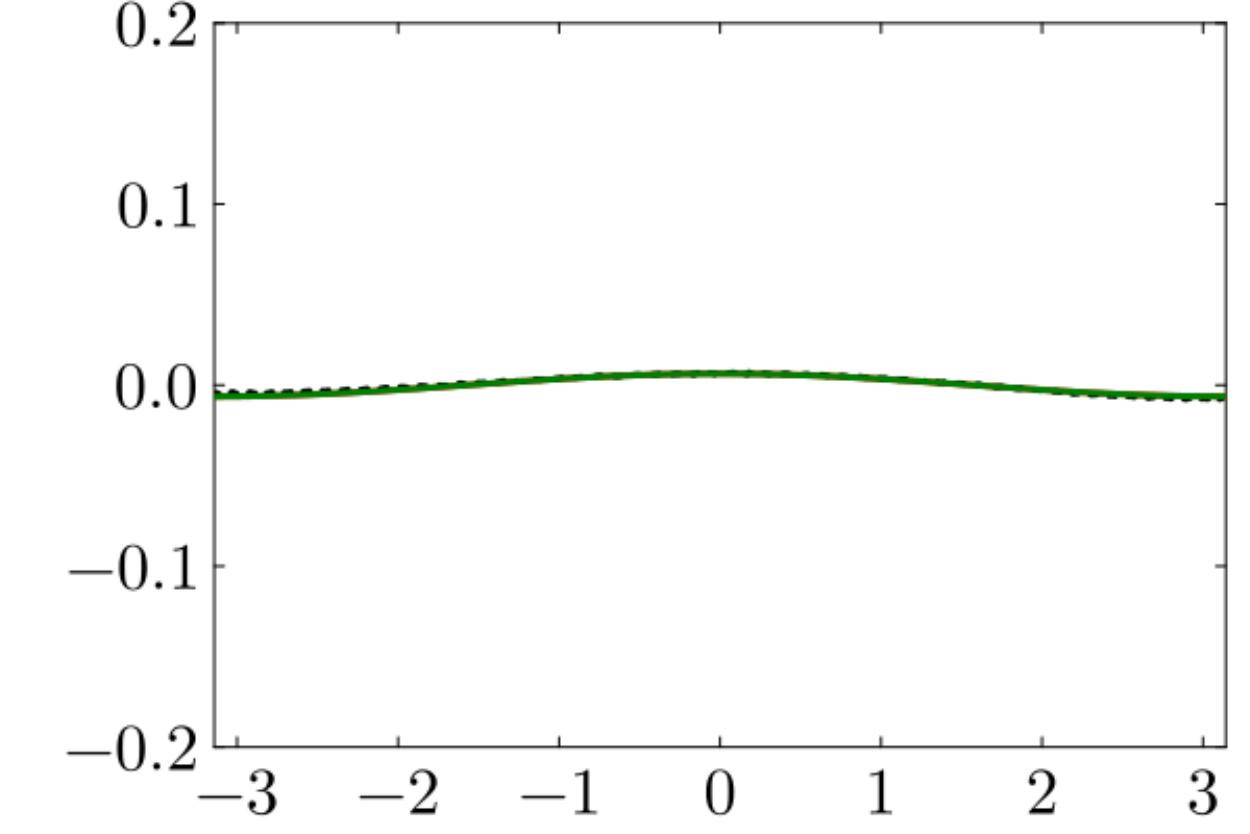}}
	\subfloat[$\dfrac{t}{T}=1$]{\includegraphics[scale=0.22]{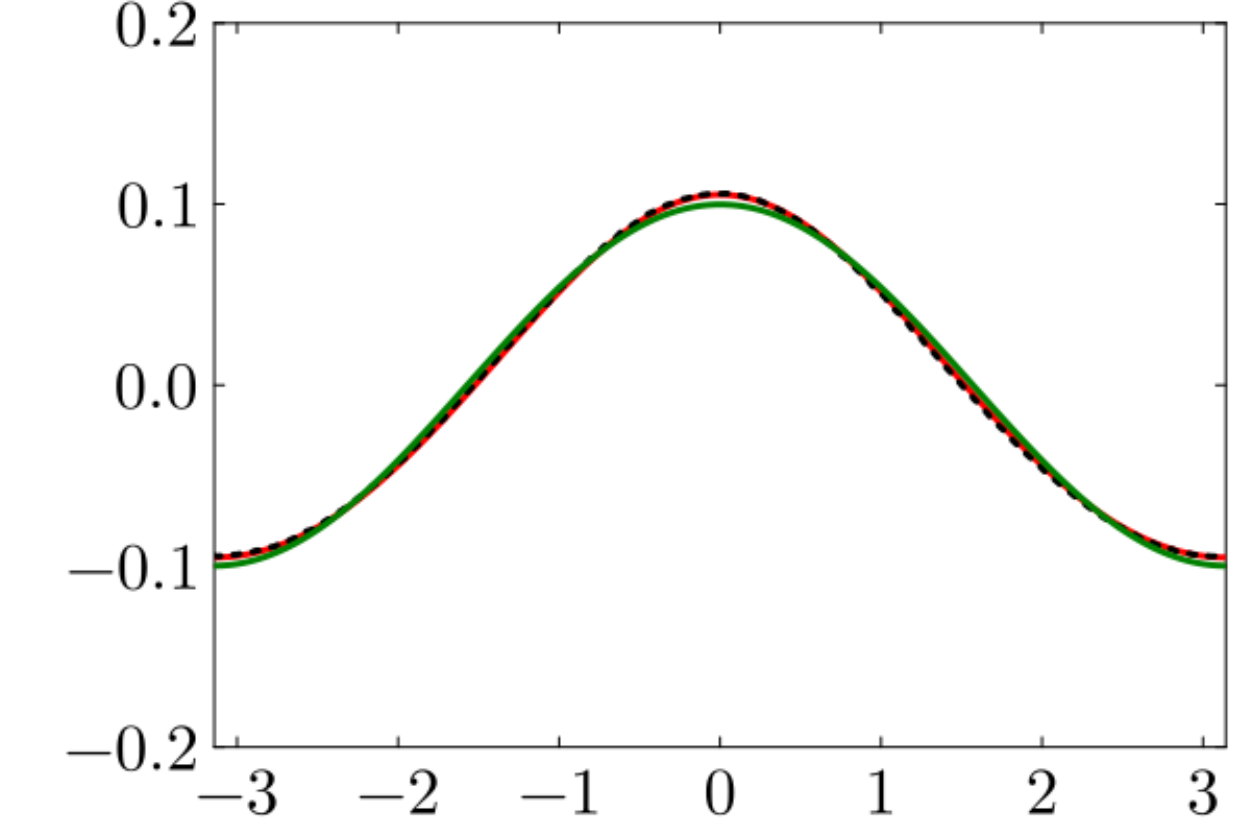}}
	\subfloat[$\dfrac{t}{T}=1.25$]{\includegraphics[scale=0.22]{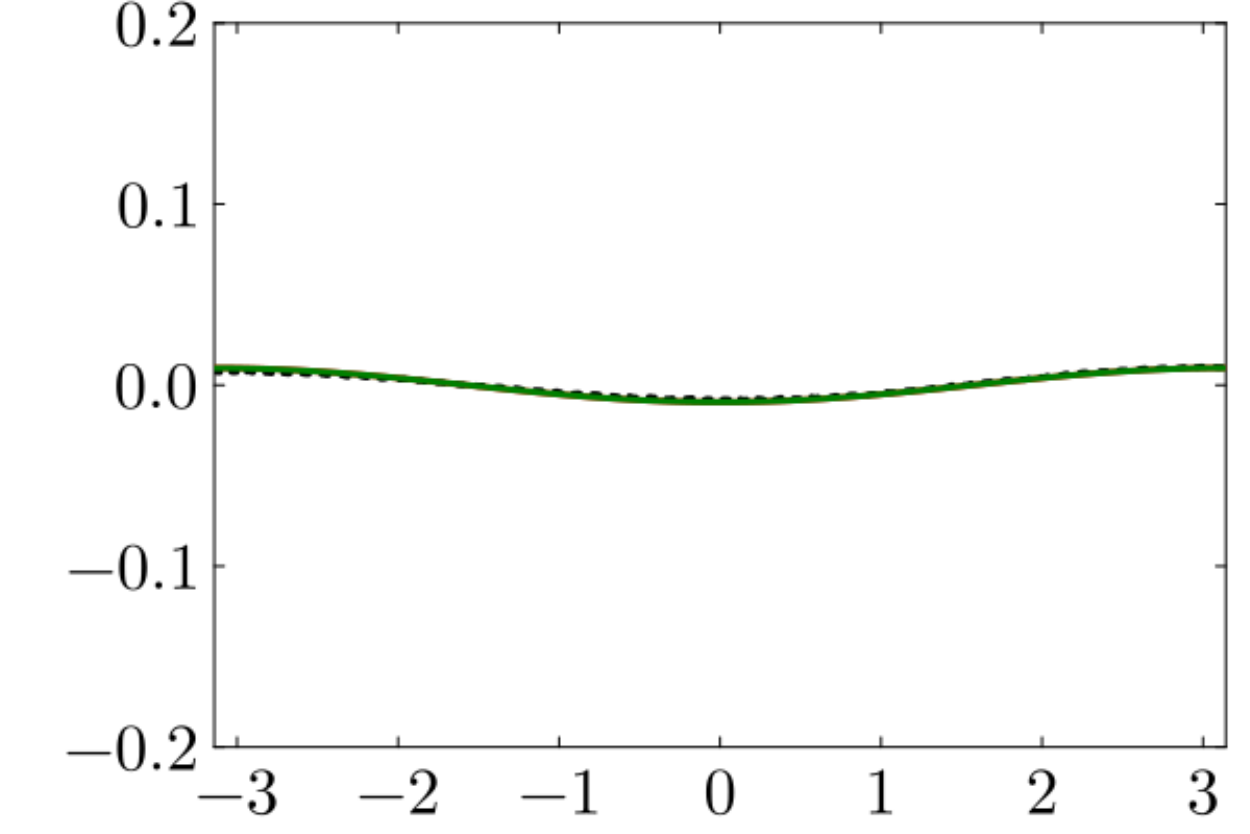}}
	\captionsetup{width=\textwidth}
	\caption{Comparison of numerical simulation using Basilisk \cite{popinet-basilisk} (labeled as Sim) initialized (see fig \ref{fig:9(a)} (Multimedia available online)) with the $\mathcal{O}(\hat{A}^5)$ accurate time-periodic \textcolor{black}{solution in eqn. (\ref{eq2}) (labeled as N)} for $\hat{A}=0.1$. \textcolor{black}{The linear approximation in eqn. (\ref{eq23}) is also plotted (L).}} 
	\label{fig:small_amp}
\end{figure*}

\begin{figure*}[htbp]
	\centering
	\subfloat[$\dfrac{t}{T}=0$]{\includegraphics[scale=0.27,trim={0.8cm 0 0 0 }]{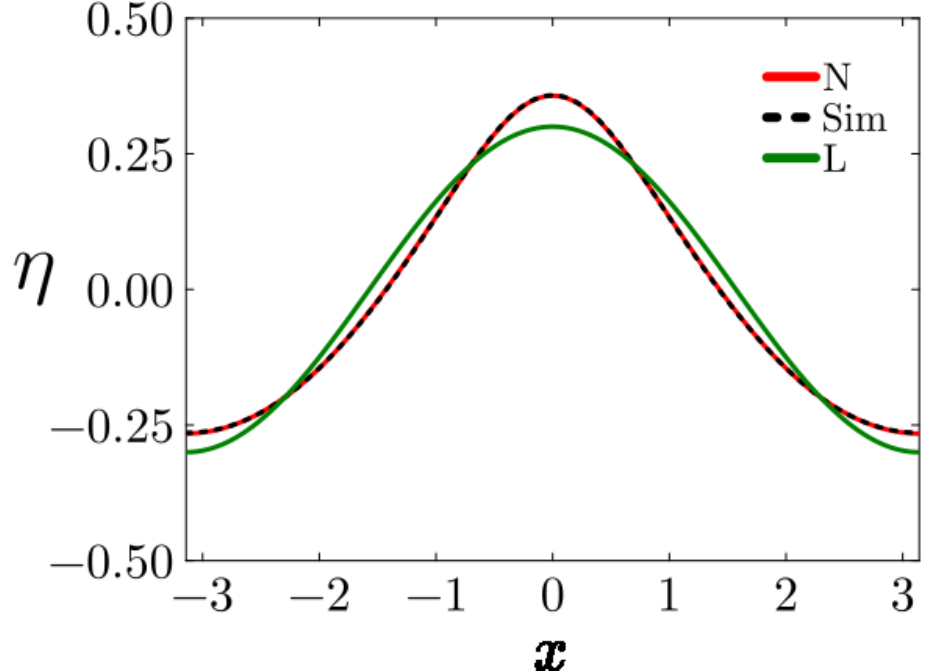}\label{fig:10(a)}}
	\subfloat[$\dfrac{t}{T}=0.25$]{\includegraphics[scale=0.22]{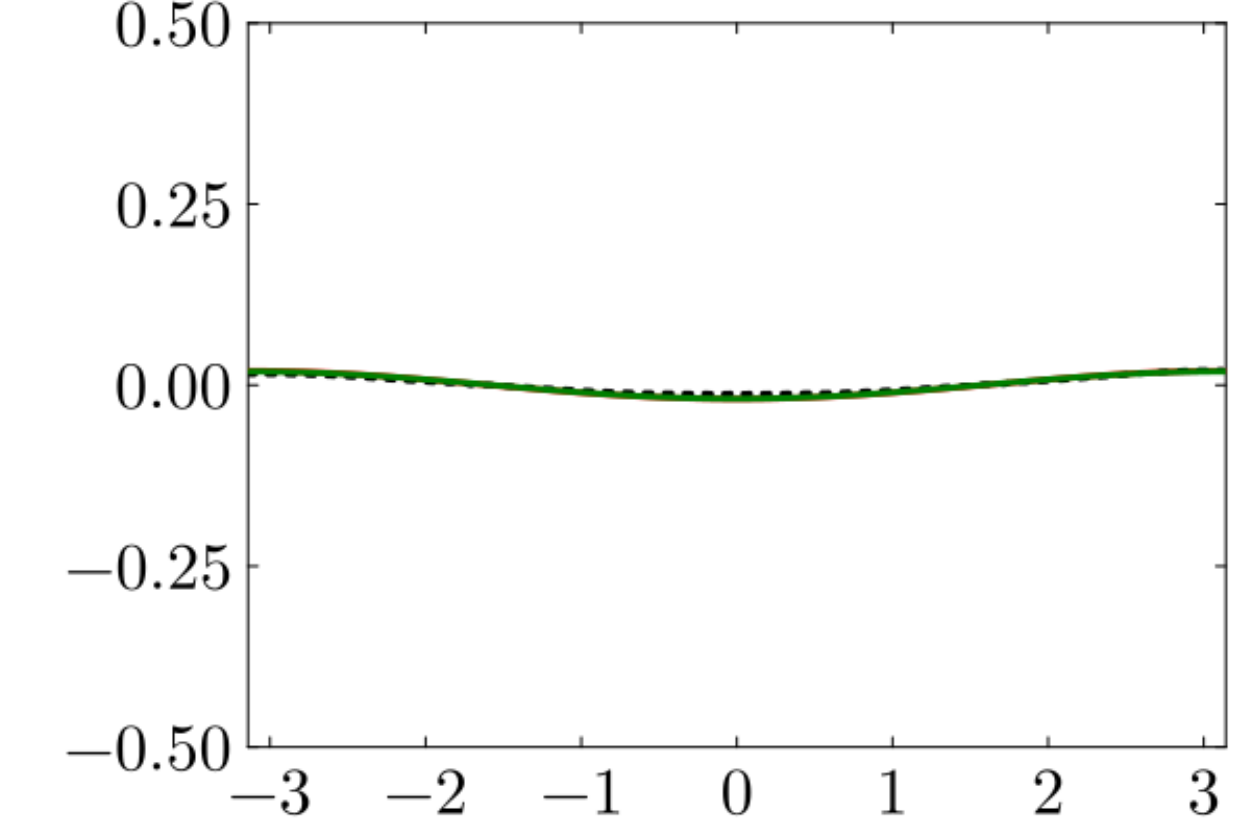}}
	\subfloat[$\dfrac{t}{T}=0.5$]{\includegraphics[scale=0.22]{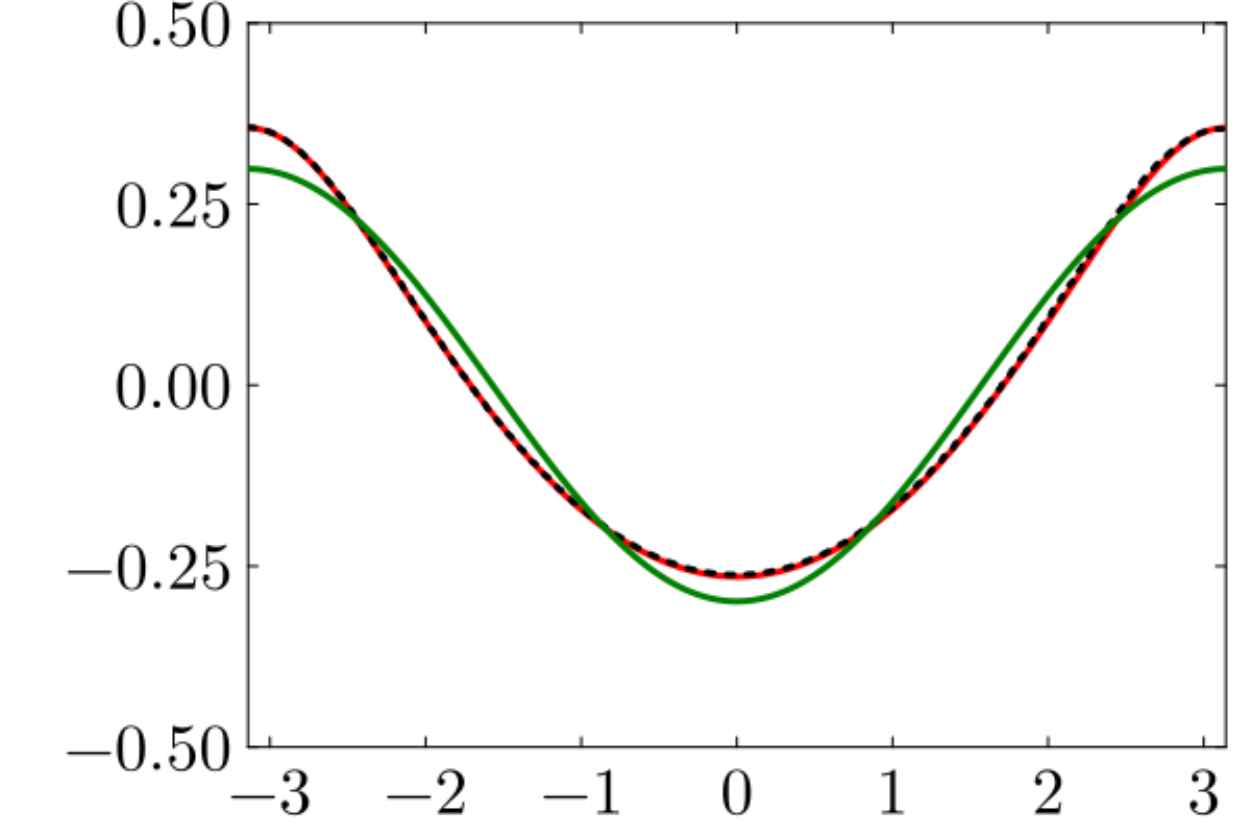}}\\
	\subfloat[$\dfrac{t}{T}=0.75$]{\includegraphics[scale=0.22,trim={0.8cm 0 0 0}]{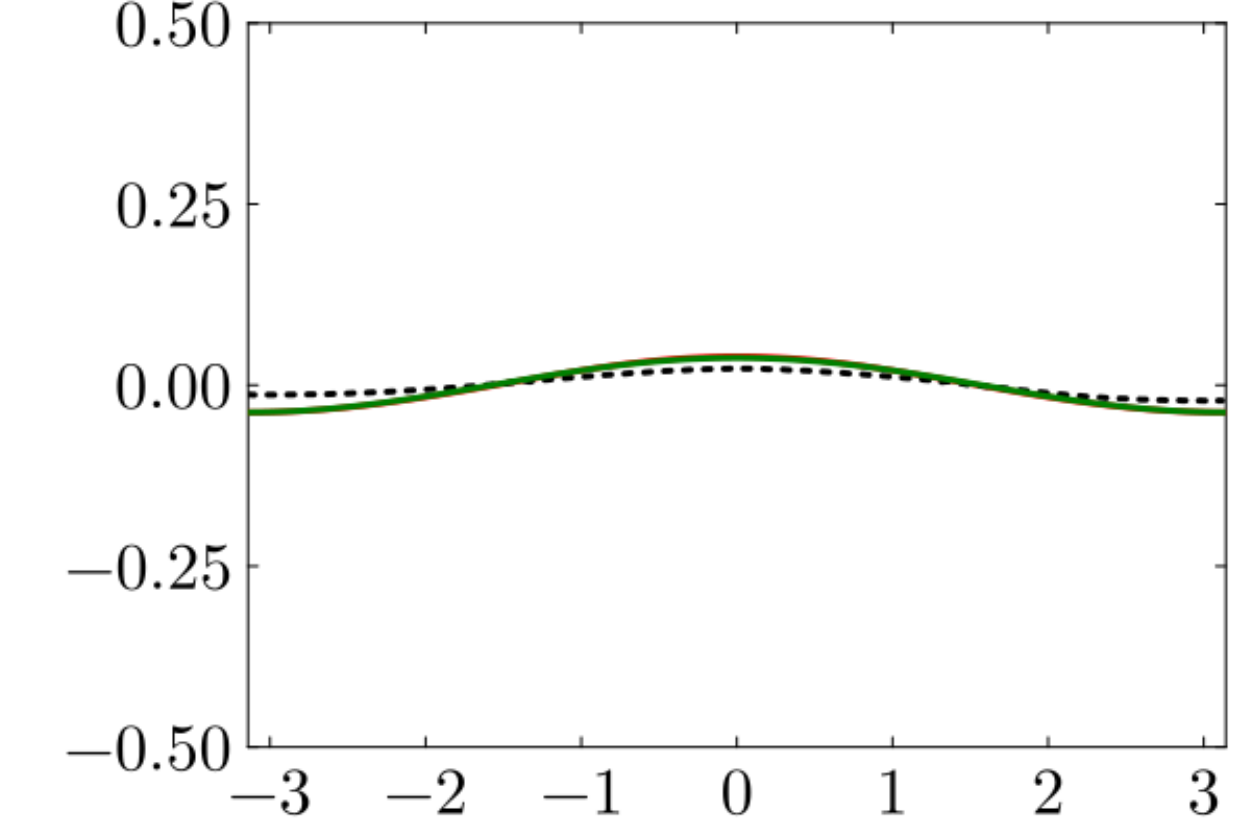}}
	\subfloat[$\dfrac{t}{T}=1$]{\includegraphics[scale=0.22]{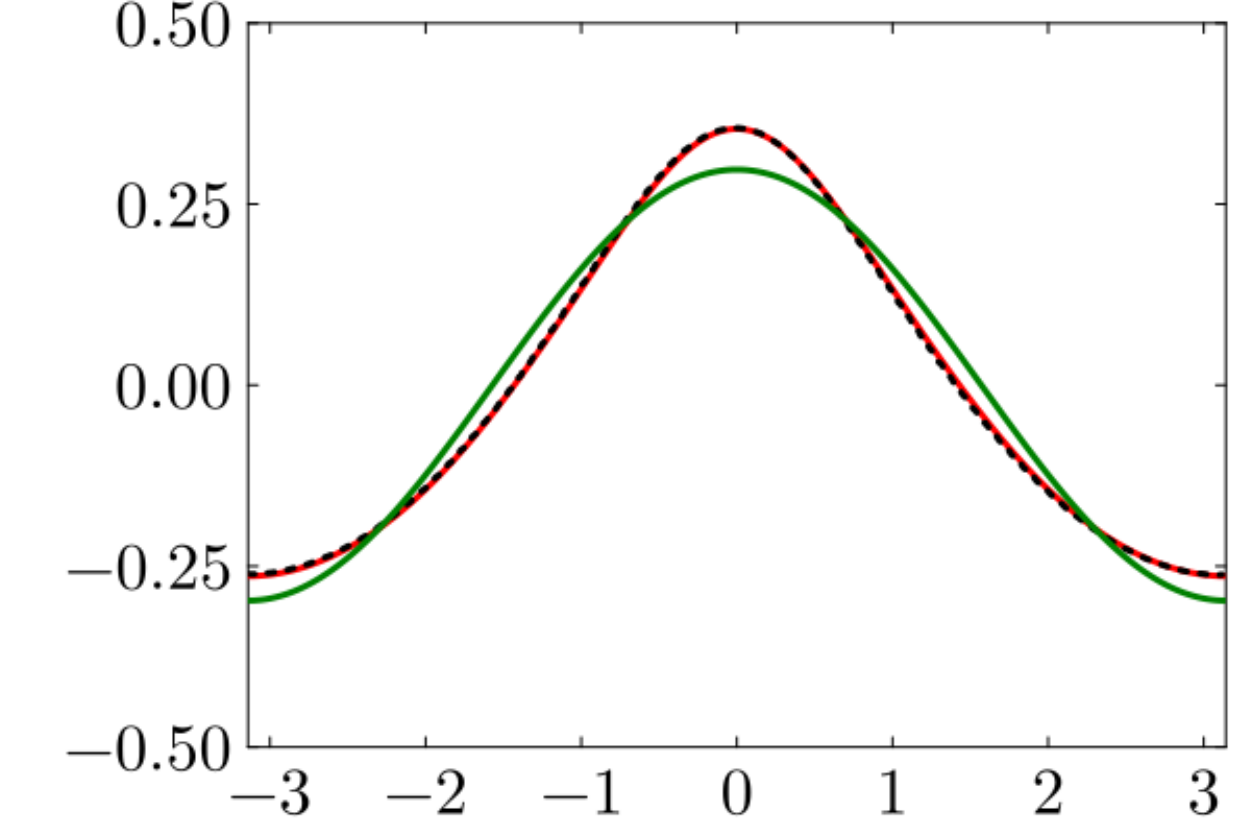}}
	\subfloat[$\dfrac{t}{T}=1.25$]{\includegraphics[scale=0.22]{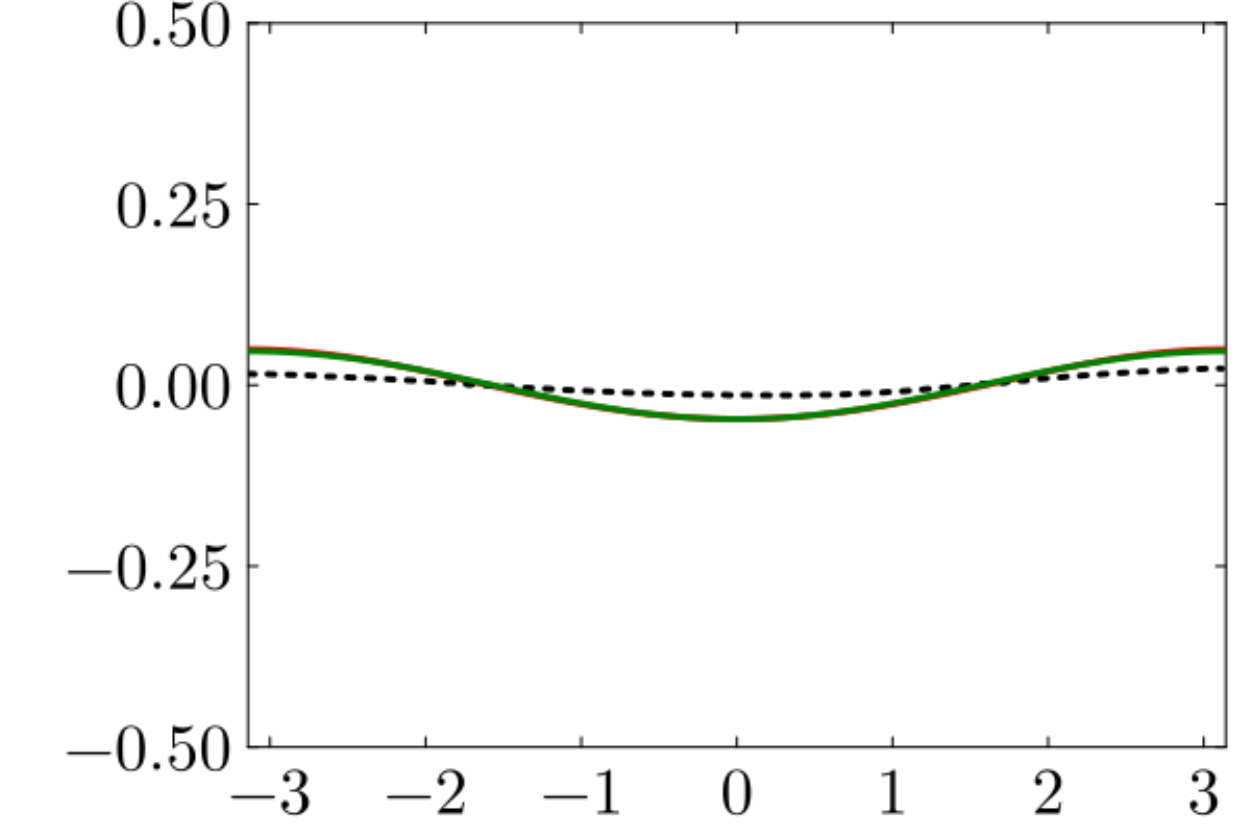}}
	\captionsetup{width=\textwidth}
	\caption{Comparison of numerical simulation (Sim, labeled as panel (a)) initialised (see fig \ref{fig:10(a)} (Multimedia available online)) with the \textcolor{black}{solution in eqn. (\ref{eq2})} for $\hat{A}=0.3$. \textcolor{black}{Curve labels are the same as in fig. \ref{fig:small_amp}.}} 
	\label{fig:med_amp}
\end{figure*}

\begin{figure*}[htbp]
	\centering
	\subfloat[$\dfrac{t}{T}=0$]{\includegraphics[scale=0.27,trim={0.8cm 0 0 0 }]{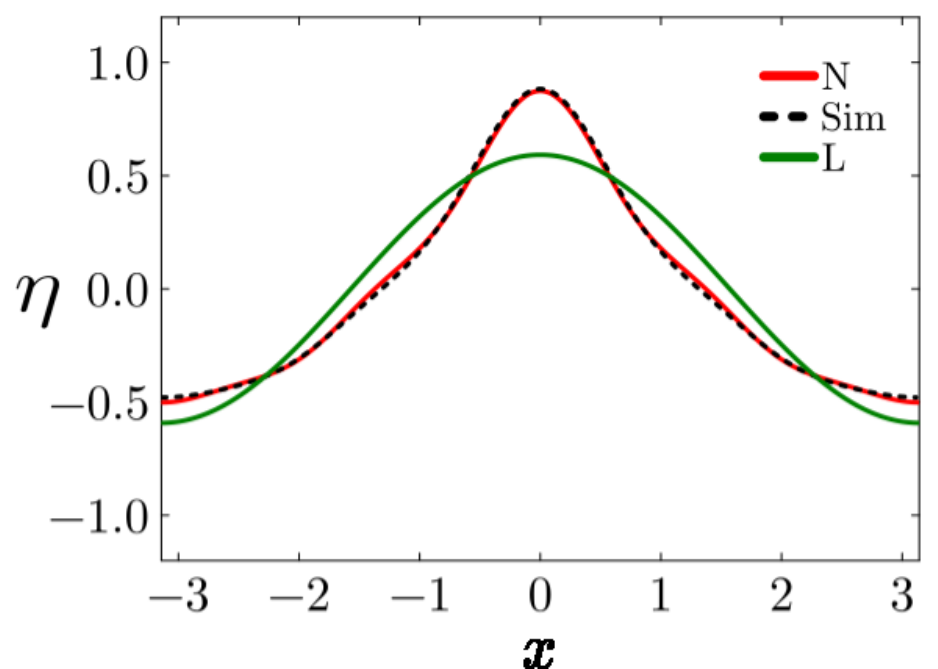}\label{fig:11(a)}}
	\subfloat[$\dfrac{t}{T}=0.25$]{\includegraphics[scale=0.22]{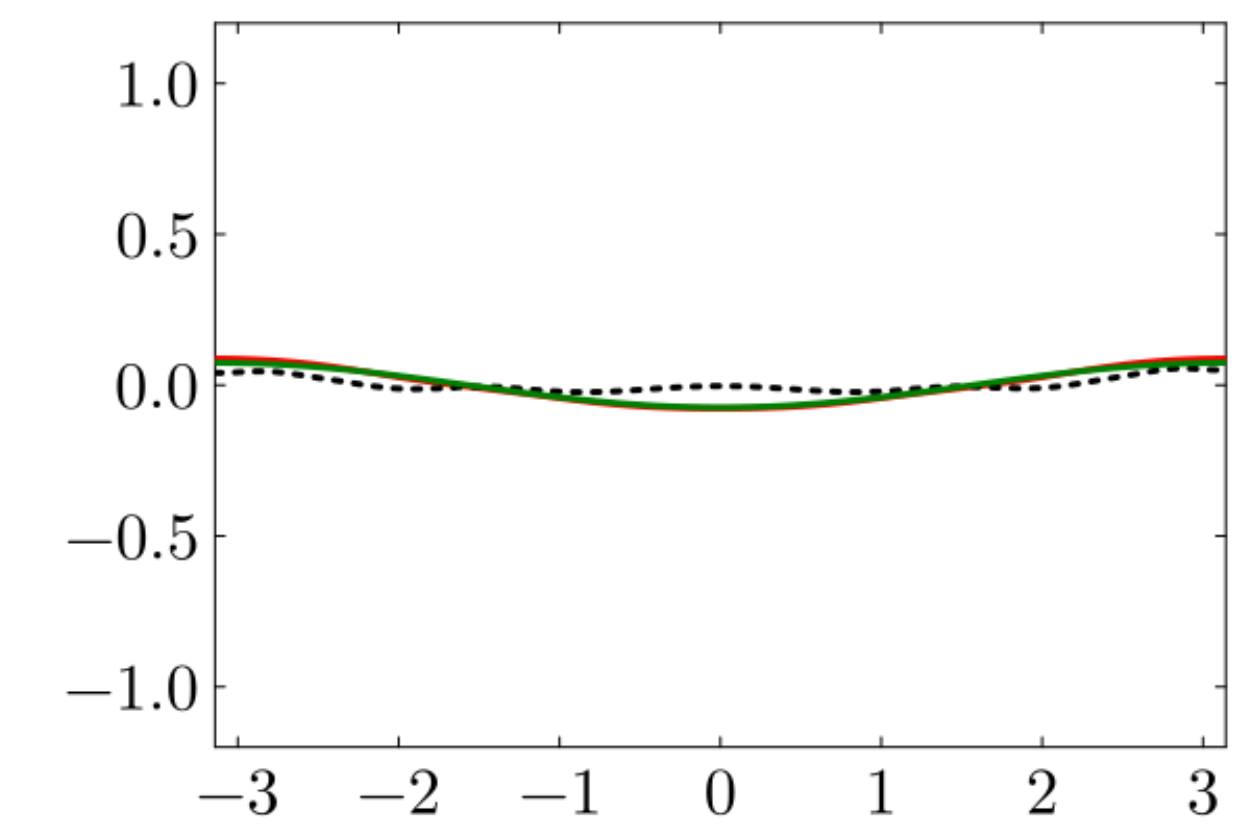}}
	\subfloat[$\dfrac{t}{T}=0.5$]{\includegraphics[scale=0.22]{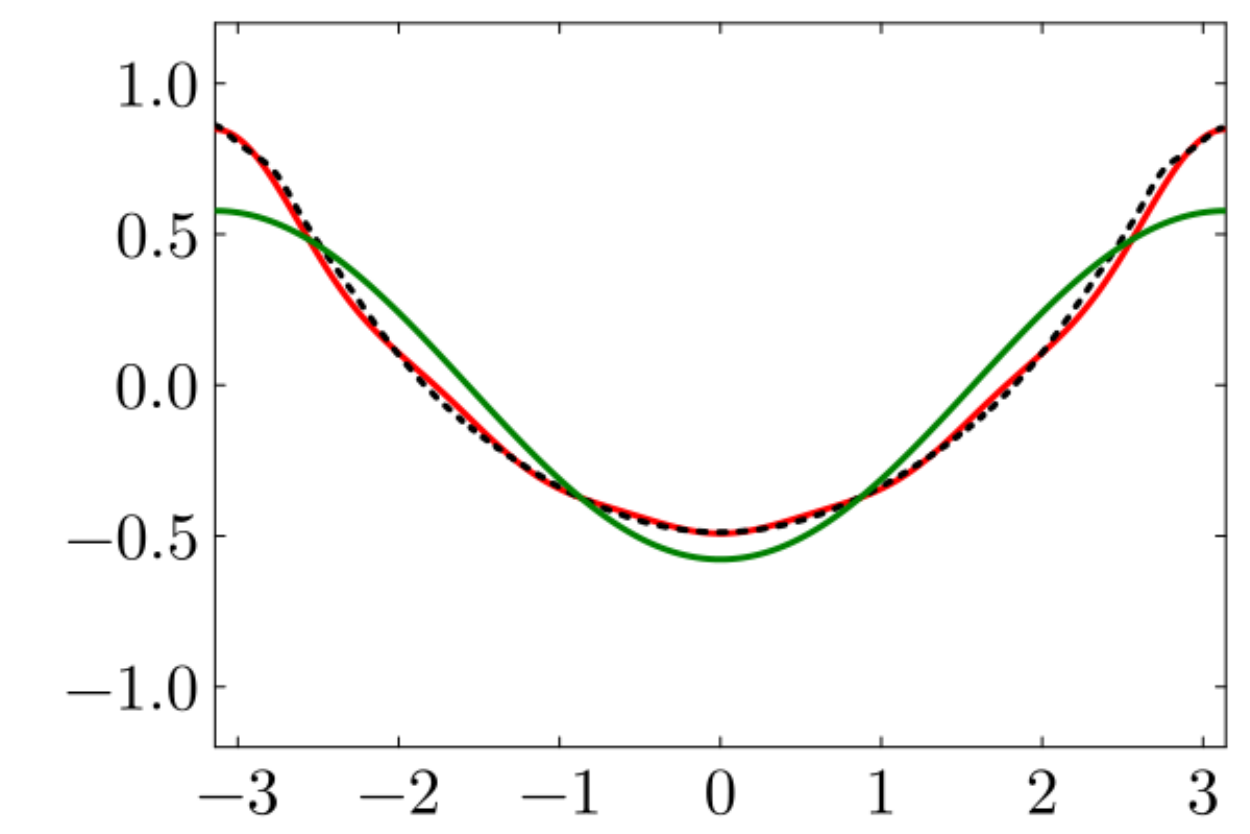}}\\
	\subfloat[$\dfrac{t}{T}=0.75$]{\includegraphics[scale=0.22,trim={0.8cm 0 0 0 }]{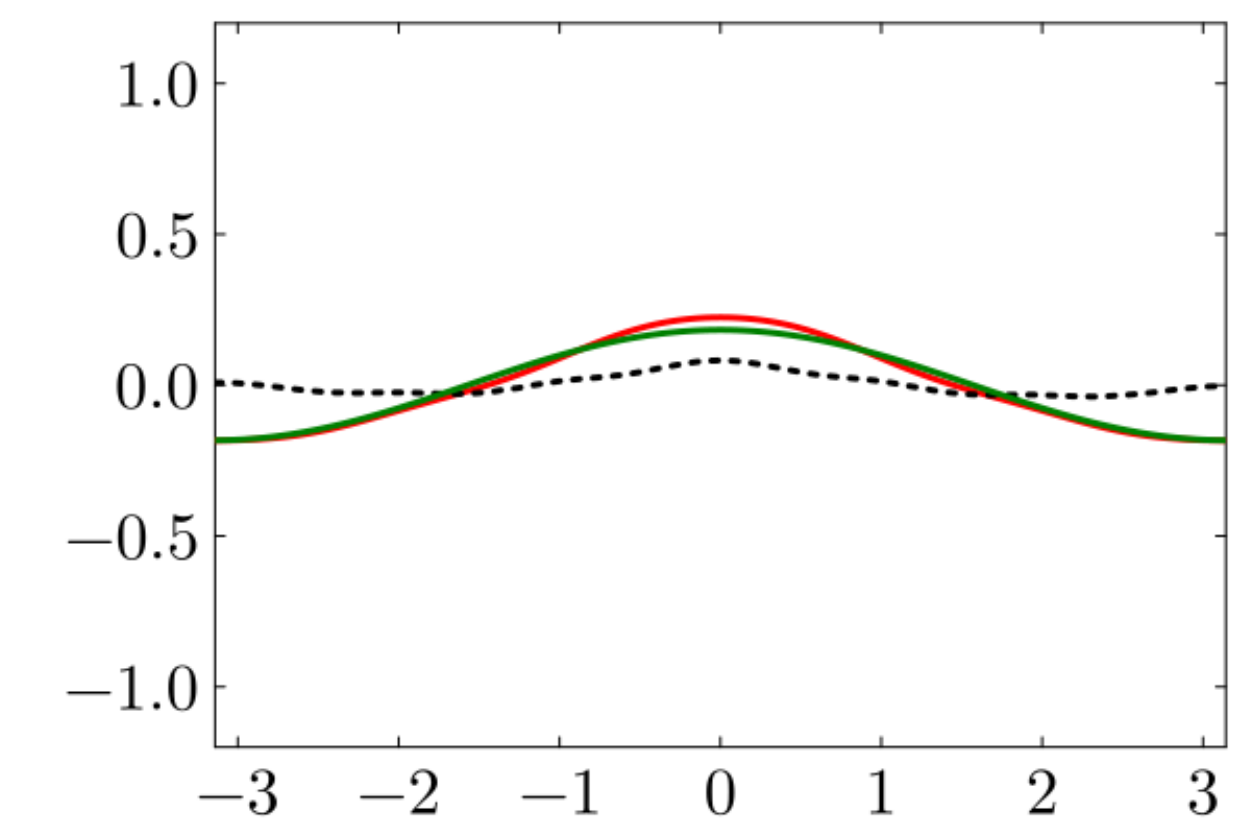}}
	\subfloat[$\dfrac{t}{T}=1$]{\includegraphics[scale=0.22]{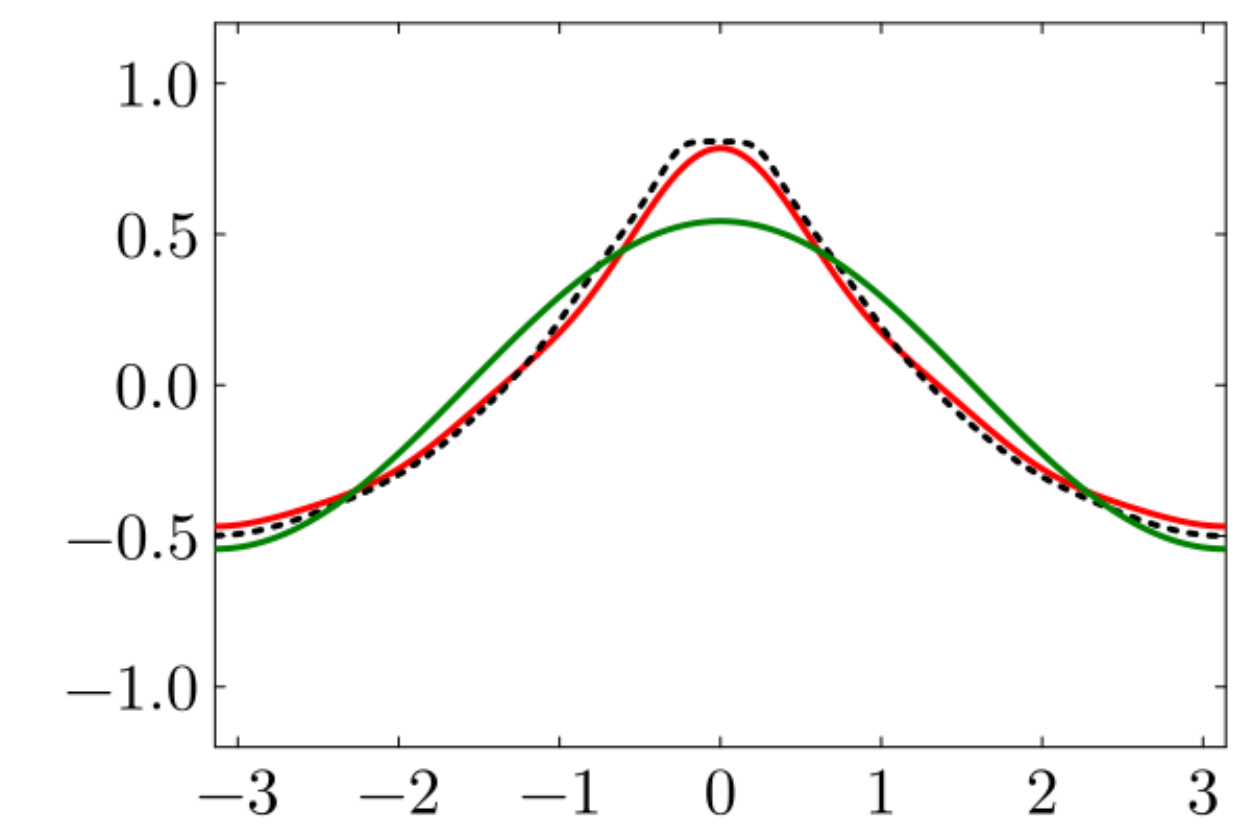}}
	\subfloat[$\dfrac{t}{T}=1.25$]{\includegraphics[scale=0.22]{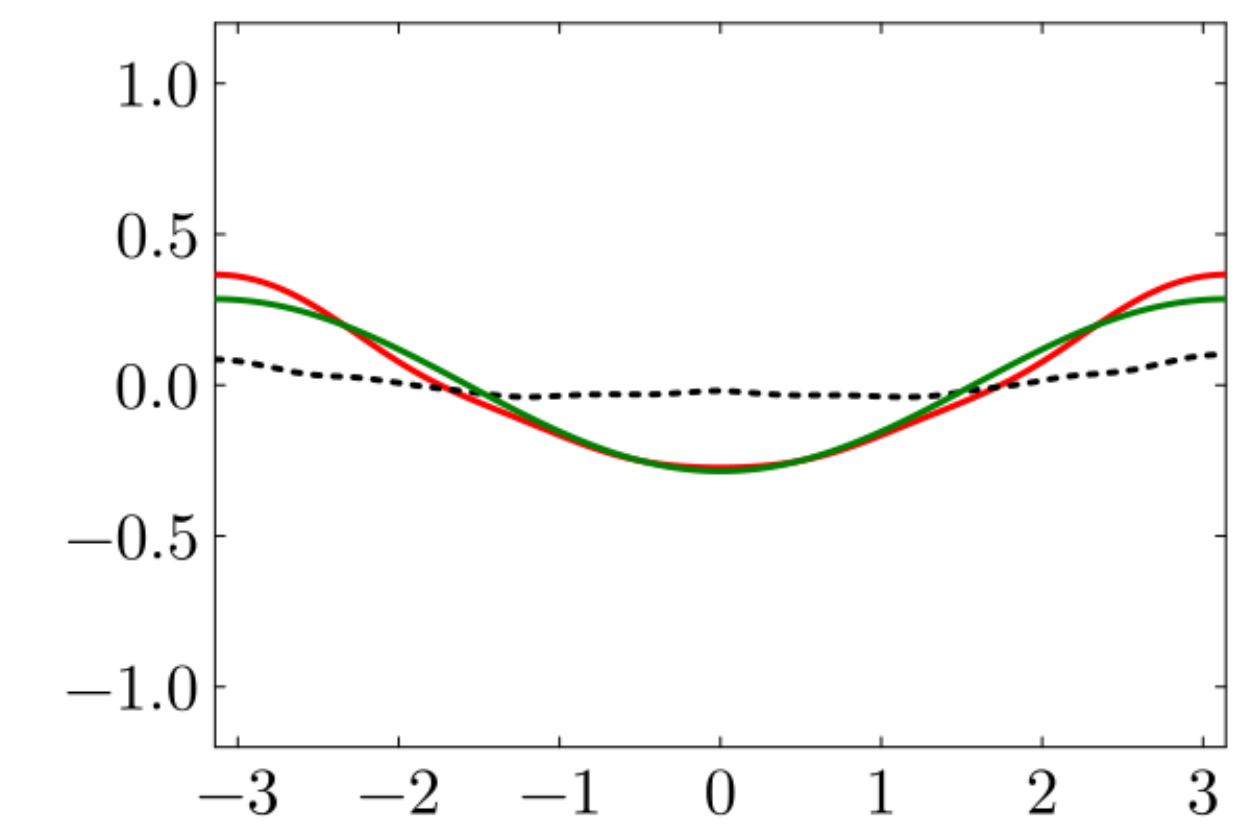}}
	\captionsetup{width=\textwidth}
	\caption{Comparison of numerical simulation(labeled as Sim) initialised (see fig \ref{fig:11(a)} (Multimedia available online)) with \textcolor{black}{the solution in eqn. (\ref{eq2})} for $\hat{A}=0.592$. \textcolor{black}{Curve labels are the same as in fig. \ref{fig:small_amp}.}} 
	\label{fig:large_amp}
\end{figure*}

\textcolor{black}{Armed with these linear stability observations, we may now return to the question posed at the end of section \ref{finiteamptheoryandsims}: \textit{why is there a difference between the simulations and the analytical expression
		in panel (b) of fig. \ref{fig:standing_waves_0_2T} at t = 8T but not in panel (a) where $\hat{A}$ is much smaller? Is there any
		possible instability at large $\hat{A}$ which could cause this ?}
Impermeable walls in our simulations restrict admissible perturbations to super-harmonics of the primary wavelength (i.e. one half, one-third, one-fourth $\ldots$ of $\lambda = 2\pi/k$), all of which are linearly stable at the steepnesses considered. It is thus likely that this distortion and the formation of pointed crests is instead related to focusing \cite{kayal2025focussing}; truncation errors in the fifth order representation of $\eta$ \cite{Penney1952} behave as numerical perturbations and can generate such pointed structures, see recent work in axisymmetric geometry \cite{kayal2025focussing,kayal2022dimples, basak2021jetting,kayal2023jet}. Finally, figs. (\ref{fig:small_amp}), (\ref{fig:med_amp}) and (\ref{fig:large_amp}) present a systematic comparison of time-snapshots of the early time evolution of the air-water interface. For reference, the linear solution in eqn. (\ref{eq23}) is also plotted. For the long-time evolution, see the multimedia available online.}

\section{Conclusion}
In this study, we have established an analogy between standing surface waves of finite-amplitude on a gas-liquid interface and oscillations of a mass attached to springs, with two degrees-of-freedom. We observe that both systems are governed by nonlinear differential equations and that these possess trivial solutions which are linearly stable. Both systems also admit exact, time-periodic solutions with a free parameter ($A$ and $\hat{A}$ respectively). The value of this determines the stability of the time-periodic solution. Stability analysis in either cases leads to Hill or Mathieu kind of equations, which dictate the short time evolution of the underlying time-periodic solution when subject to perturbations. \textcolor{black}{For the mechanical oscillator, the stable and unstable regimes can be discerned from the corresponding stability charts. For the interfacial oscillator subjected to a super-harmonic perturbation the governing equation for the time-dependent amplitude of perturbation is shown to be a  \textit{novel Mathieu-like} equation (eqn. (\ref{eq33}))}, in a reduced order description. This equation predicts only stable oscillations. In our computations for the interfacial oscillator, we also observe that the interface when initialised in the form of a steep, time-periodic, stable shape \cite{Penney1952}, does not preserve its shape at sufficiently large time. We argue that this behaviour is not due to an underlying linear instability and hypothesize the role of truncation errors in representing the base-state accurately, in generating this. \textcolor{black}{This argument is further reinforced in fig. \ref{fig:mathieulike} by the numerical solution of the \textit{Mathieu-like} equation that is shown to govern the time-dependent amplitude of a super-harmonic perturbation.}

In conclusion, we emphasize that despite the analogy between the two systems presented here, there remain important differences. The spring-mass system has only two degrees of freedom. This reflects in the fact that the `shape' of its time-periodic solution may be expressed as $\left[x(t)\;y(t)\right]^{T}=A\left[1\quad 0\right]^{T}\cos(\omega_0t)$. In contrast, the interfacial oscillator has a far more complex `shape' reflected in the expression $\eta(x,t;\hat{A})$ in eqn. (\ref{eq2}); crucially the shape of this time-periodic solution depends on the free parameter $\hat{A}$. Moreover, the base-state frequency of the interfacial oscillator also depends on $\hat{A}$ (refer description of eqn. (\ref{eq2}), unlike that of the frequency of the spring-mass system $\omega_0=\sqrt{2k/m}$ which is independent of $A$. The interfacial oscillator admits a countably infinite set of eigenmodes, as discussed in section \ref{sec:stabilityInterfacial},  leading to rich behaviour of coalescence of modes and thereby instability; the spring-mass system in contrast, has only two modes. \textcolor{black}{Also, the novel eqn. (\ref{eq33}) derived for time-dependent amplitude of super-harmonic perturbation in interfacial oscillator is a \textit{Mathieu-like} and not an exact \text{Mathieu} equation as in case of mechanical oscillator. As discussed in section \ref{sec:stabilityInterfacial}, we expect better stability predictions by using more accurate representation of the base-state and perturbations. This may render coefficient of $a'(t)$ in eqn. (\ref{eq33}) to zero. However, we acknowledge that this may lead to tedious algebra and extremely lengthy equations. } For the interested reader, there are several other examples of finite-amplitude, natural oscillations of standing or travelling type encountered in situations of multiphase flow (see \citet{Stokes_2009} or \citet{Tsamopoulos_Brown_1983}).

\backmatter

\bmhead{Supplementary information}
Multimedia(video) files related to figs. \ref{fig:small_amp}, \ref{fig:med_amp} and \ref{fig:large_amp} are available in mp4 format. The file names corresponding to each figure are same as that of figure numbers.

\bmhead{Acknowledgements}

	We thank the organization of the International Research Network (IRN) under the auspice of CNRS. NY is supported by the Prime Minister Research Fellowship (PMRF) by Ministry of Education, Govt of India and gratefully acknowledges this support. We thank IIT Bombay, Seed funding for collaboration \& partnership projects (SCPP) scheme (Institute of Eminence, IIT Bombay) which seeded the collaboration between RD and SA. We gratefully acknowledge financial support from DST-SERB (Govt. of
India) grants MTR/2019/001240, CRG/2020/003707, SPR/2021/000536 and Ministry of Education (Govt. of India) MoE-STARS/STARS-2/2023-0595, for supporting research on surface waves and their breaking.

\section*{Statements and Declarations}

\textbf{Conflict of interest}
The authors declare no competing interests.

\begin{appendices}

\section{: Derivation of equations (\ref{eq3}) and (\ref{eq4})}\label{app1}

We derive the equations of motion (\ref{eq3}) and (\ref{eq4}) of the system presented in figure \ref{fig1_springmass}. The Lagrangian $\mathcal{L}$ of the system is defined as:
\begin{align}
	\mathcal{L} = T- V
	= \dfrac{1}{2}m\left(\bm{\dot}{x}^2 + \bm{\dot}{y}^2\right)-\dfrac{1}{2}k\left(\sqrt{(a+x)^2 + y^2}-L\right)^2 - \dfrac{1}{2}k\left(\sqrt{(a-x)^2+ y^2}-L\right)^2 \label{a1}
\end{align}
The governing equations of motion of this system follow then from the Euler-Lagrange equations of motion. These are:

\begin{align}
	&\dfrac{d}{dt}\left( \dfrac{\partial \mathcal{L}}{\partial \bm{\dot}{x}}\right) = \left(\dfrac{\partial \mathcal{L}}{\partial x} \right) \nonumber \\
	&\Rightarrow m\ddot{x} + k \left( 1 - \dfrac{L}{\sqrt{(a+x)^2 + y^2}} \right)(a+x)  - k\left(1- \dfrac{L}{\sqrt{(a-x)^2 + y^2}}  \right)(a-x)=0 \label{a2}\\
	&\dfrac{d}{dt}\left( \dfrac{\partial \mathcal{L}}{\partial \bm{\dot}{y}}\right)= \left(\dfrac{\partial \mathcal{L}}{\partial y} \right)  \nonumber \\
	&\Rightarrow m\bm{\ddot}{y} + k \left( 1 - \dfrac{L}{\sqrt{(a+x)^2 + y^2}} \right)y  + k\left(1- \dfrac{L}{\sqrt{(a-x)^2 + y^2}}  \right)y=0 \label{a3}    
\end{align} 

Linearisation about the trivial solution $x(t)=0, y(t)=0$ using the expansion $x(t)=0+\delta x(t),\;y(t)=0+\delta y(t)$ yields the equations,

\begin{align}
	&\delta \ddot{x} + \dfrac{k}{m} \left( 1 - \dfrac{L}{\sqrt{(a+\delta x)^2 + \delta y^2}} \right)(a+\delta x)  - \dfrac{k}{m}\left(1- \dfrac{L}{\sqrt{(a-\delta x)^2 + \delta y^2}}  \right)(a-\delta x)=0 \label{a4}\\
	&\delta \ddot{y}  + \dfrac{k}{m} \left( 1 - \dfrac{L}{\sqrt{(a+\delta x)^2 + \delta y^2}} \right)\delta y + \dfrac{k}{m}\left(1- \dfrac{L}{\sqrt{(a-\delta x)^2 + \delta y^2}}  \right)\delta y=0 \label{a5}
\end{align}
Retaining only upto linear order in perturbation variables, we obtain from eq. \ref{a4}
\begin{align}
	\delta \ddot{x}   + \omega_0^2 \delta x = 0 \label{a6}
\end{align}
A similar process with equation (\ref{a5}) leads to :
\begin{align}
	\delta \ddot{y} + \omega_0^2 \left(1 - \dfrac{L}{a}\right)\delta y = 0 \label{a7}
\end{align}
For the time-periodic solution, considering the perturbation of base-state  $x= x_b(t) + u(t) = A \cos(\omega_0 t) + u(t), \; y= y_b(t) + v(t) = 0+v(t)$. Substituting this in eqn. (\ref{a4}), and linearising while recalling that $\omega_0^2 =\dfrac{2k}{m}$, we obtain
\begin{align}
	\bm{\ddot}{u} + \omega_0^2 u =0 \label{a8}
\end{align}
For eqn. (\ref{a5}), using the same expansion $x=A \cos(\omega_0 t) + u(t), \; y=0+v(t)$ we obtain
\begin{align}
	\bm{\ddot}{v} + \dfrac{2k}{m} \left [1 - \dfrac{L}{a}\left( \dfrac{1}{1 - \left(\left(\dfrac{A}{a} \right)\cos(\omega_0 t) + \dfrac{u}{a} \right)^2 }\right) \right]v(t) = 0 \label{a9}
\end{align}
Eqn. (\ref{a9})  can be re-written as Hill's differential equation (\ref{eq8}) by linearisation or as the Mathieu equation (\ref{eq9}), by retaining terms upto $\mathcal{O}(A^2)$. This algebra is quite easy and is not provided here.

\section{: Floquet analysis on the Hill and Mathieu equations}\label{Floquet}

Consider the Mathieu equation (\ref{eq11}) which can be written as set of first order ODEs by assuming $\dfrac{dv}{d\tau} \equiv \theta(\tau)$, thus resulting in the following :
\begin{align}\label{b1}
	\dfrac{dv}{d\tau}= \theta(\tau), \; \dfrac{d\theta}{d\tau} = -\left(\delta -2\epsilon \cos(\tau) \right)v(\tau)
\end{align}
The coefficients of the Mathieu equation (\ref{eq9}) are $2\pi$ periodic in $\tau$. We can numerically solve equation (\ref{b1}) with the  initial conditions $v(\tau = 0)=1 , \theta(\tau=0)=0$ and $v(\tau = 0)=0 , \theta(\tau=0)=1$ respectively, to obtain two solution trajectories that are linearly independent. The ODEs are numerically solved for time span $\tau \in [0, 2\pi]$ using Dormand-Prince  Runge-Kutta method (same as ode45 in MATLAB) using DifferentialEquations.jl \cite{rackauckas2017differentialequations}, with an absolute and relative tolerance of $10^{-16}$. Lets assign $v_1 (\tau), \theta_1  (\tau)$ and $v_2 (\tau), \theta_2 (\tau)$ as notation to the trajectories obtained by solving them numerically.
\begin{align}
	M = \begin{bmatrix}
		\theta_1 (\tau=2\pi) & \theta_2 (\tau=2\pi) \\
		v_1 (\tau = 2\pi) & v_2 (\tau=2\pi)
	\end{bmatrix}
\end{align}
Depending on the eigenvalues $\lambda$ of the monodromy matrix $M$ (see \citet{KovacicRand}), we can classify whether the solution for equation ($\ref{eq9}$) is stable or unstable \cite{Rand}
\begin{align}
	\lambda^2 - \text{tr}(M)\lambda + \text{det}(M) =0 \\
	\lambda_1 , \lambda_2 = \dfrac{\text{tr}(M) \pm \sqrt{\text{tr}(M)^2-1}}{2}
\end{align}
where \say{$\text{tr}$} and \say{$\text{det}$} refer to trace and determinant respectively. Thus, if either of the eigenvalues ($\lambda_1, \lambda_2$) has modulus greater than $1$, there is instability.
\nocite{*}
Similar steps are followed for the Hill equation (\ref{eq8}) re-written as a set of first order ODEs (similar to eqn (\ref{b1})) as follows:
\begin{align}
	\dot{v} = \theta(t), \;	\dot{\theta} = -\omega_0^{2}\left[1 -  \dfrac{ \dfrac{L}{a}}{{1 - \left(\dfrac{A}{a}\right)^2 \cos^{2}(\omega_0 t) }} \right]v(t) \label{b5}
\end{align}
The exact same steps are followed to obtain the monodromy matrix $M$ and its eigenvalues $\lambda_1, \lambda_2$ as discussed earlier for Mathieu equation.

\section{\textcolor{black}{: Coefficients for series representation in eqn. \ref{eq2} in finite-amplitude standing wave}\label{app4}}

\textcolor{black}{\begin{flalign}
	&b_1 = \left( \hat{A} + \dfrac{3}{2} \hat{A}^3 - \dfrac{137}{3072} \hat{A}^5 \right) \sin\left(\omega_0 t\right) + \left( \dfrac{1}{16} \hat{A}^3 - \dfrac{11}{5376} \hat{A}^5 \right) \sin\left(3\omega_0 t\right) + \dfrac{163}{21504} \hat{A}^5 \sin\left(5\omega_0 t\right), \nonumber \\
	&b_2 = \dfrac{1}{4} \hat{A}^2 + \dfrac{1}{16} \hat{A}^4 - \left( \dfrac{1}{4} \hat{A}^2 - \dfrac{25}{192} \hat{A}^4 \right)\cos \left(2\omega_0 t\right)- \dfrac{67}{1344} \hat{A}^4 \cos\left(4\omega_0 t\right), \nonumber \\
	&b_3 = \left( \dfrac{9}{32} \hat{A}^3 - \dfrac{1}{256} \hat{A}^5 \right) \sin\left(\omega_0 t\right) - \left( \dfrac{3}{32} \hat{A}^3 - \dfrac{2195}{14336} \hat{A}^5 \right) \sin\left(3\omega_0 t\right)  - \dfrac{16365}{473088} \hat{A}^5 \sin\left(5\omega_0 t\right),\nonumber\\
	&b_4 = \dfrac{1}{8} \hat{A}^4 - \dfrac{1}{6} \hat{A}^4 \cos\left(2\omega_0 t\right) + \dfrac{1}{24} \hat{A}^4 \cos\left(4\omega_0 t\right), \nonumber\\
	&b_5 = \dfrac{145}{768} \hat{A}^5 \sin\left(\omega_0 t\right) - \dfrac{5}{3072} \hat{A}^5 \sin\left(3\omega_0 t\right) +\dfrac{8}{3072} \hat{A}^5 \sin\left(5\omega_0 t\right). \nonumber
\end{flalign}
\noindent where $\omega_0^2  \equiv gk\left( 1 - \dfrac{\hat{A}^2}{4} - \dfrac{13}{128} \hat{A}^4\right) $, $b_0 = 0$ (since we choose $y=0$ along the mean level of water in fig. \ref{fig0_SurfaceWave}). These coefficients were derived by \citet{Penney1952}.}

\section{: Linear stability analysis of finite-amplitude, standing wave}\label{app3}
Considering the perturbation of 
	the finite-amplitude, standing wave represented by $\eta_b(x,t)$ and $\phi_b(x,y,t)$ (in section \ref{sec:stabilityInterfacial}) the equations  for $y_s$ and $\Phi$ are as follows : 
	\begin{align}
		&y_s(x,t) = \eta_b(x,t) + \epsilon \tilde{\eta}(x,t) \label{eqc1} \\
		&\Phi(x,y,t) = \phi_b (x,y,t) + \epsilon \tilde{\phi}(x,y,t) \label{eqc2}
	\end{align}
	Taylor expanding equations (\ref{eq15}), (\ref{eq16}) about the base state $\eta_b, \phi_b$ leads to : (at $y=\eta_b(x,t)$):
	\begin{align}
		& \dfrac{\partial y_s}{\partial t} + \epsilon \tilde{\eta} \cancelto{0}{\dfrac{\partial}{\partial z}\left(\dfrac{\partial y_s}{\partial t}\right)} + \left(\dfrac{\partial y_s}{\partial x}\right)\left[\dfrac{\partial \Phi}{\partial x} + \epsilon \tilde{\eta} \dfrac{\partial}{\partial y}\left(\dfrac{\partial \Phi}{\partial x}\right) \right]  = \dfrac{\partial \Phi}{\partial y} + \epsilon \tilde{\eta} \dfrac{\partial^2 \Phi}{\partial y^2} \label{eqc3} \\ 
		& \dfrac{\partial \Phi}{\partial t} + \epsilon \tilde{\eta} \dfrac{\partial}{\partial y}\left(\dfrac{\partial \Phi}{\partial t} \right) + \dfrac{1}{2}\left[|\nabla \Phi|^2 + \epsilon \tilde{\eta}\dfrac{\partial}{\partial y}|\nabla \Phi|^2 \right] +  g(\eta_b + \epsilon \tilde{\eta}) = 0 \label{eqc4}
	\end{align}
	Equations (\ref{eqc3}) can be further expanded upto $O(\epsilon)$ as follows :
	\begin{align}
		&\dfrac{\partial \eta_b}{\partial t} + \epsilon \dfrac{\partial \tilde{\eta}}{\partial t} + \left(\dfrac{\partial \eta_b}{\partial x} + \epsilon \dfrac{\partial \tilde{\eta}}{\partial x}\right)\Biggl[\dfrac{\partial \phi_b}{\partial x} + \epsilon \dfrac{\partial \tilde{\phi}}{\partial x} + \epsilon \tilde{\eta} \dfrac{\partial}{\partial y}\left(\dfrac{\partial \phi_b}{\partial x} + \right. \left.\epsilon \dfrac{\partial \tilde{\phi}}{\partial x}\right ) \Biggr]\nonumber\\
		&\quad = \dfrac{\partial \phi_b}{\partial y}  + \epsilon \dfrac{\partial \tilde{\phi}}{\partial y} + \epsilon \tilde{\eta} \left[\dfrac{\partial^2 \phi_b}{\partial y^2} + \epsilon \dfrac{\partial^2 \tilde{\phi}}{\partial y^2}  \right] \label{eqc5} 
	\end{align} 
	Note that $O(1)$ terms are the exact equations (\ref{eq15}-\ref{eq16}) and are satisfied by the base-state, and thus can be omitted.
	Retaining only up to $O(\epsilon A^2)$ , and collecting the terms, yields the following equation :
	\begin{align}
		&\dfrac{\partial \tilde{\eta}}{\partial t} + \Biggl(\dfrac{\partial \phi_b}{\partial x}\dfrac{\partial \tilde{\eta}}{\partial x} + \dfrac{\partial \tilde{\phi}}{\partial x}\dfrac{\partial \eta}{\partial x} + \tilde{\eta}\Bigg[\dfrac{\partial \eta_b}{\partial x}\dfrac{\partial}{\partial y}\left(\dfrac{\partial \phi_b}{\partial x}\right)\Bigg] -  \dfrac{\partial \tilde{\phi}}{\partial y}-\tilde{\eta}\dfrac{\partial^2 \phi_b}{\partial y^2}  \Biggr)=0 \label{eqc6}
	\end{align}
	Similarly, we obtain the following equation by performing similar steps (eqn (\ref{eqc3}) to (\ref{eqc6})). Expanding (\ref{eqc4}) upto $O(\epsilon)$  can be written as follows:
	\begin{align}
		&\dfrac{\partial \phi_b}{\partial t} + \epsilon \dfrac{\partial \tilde{\phi}}{\partial t} + \epsilon \tilde{\eta}\dfrac{\partial}{\partial y}\Bigg[\dfrac{\partial \phi_b}{\partial t}+\epsilon \dfrac{\partial \tilde{\phi}}{\partial t}\Bigg] + \dfrac{1}{2}\Biggl(|\nabla (\phi_b + \epsilon \tilde{\phi})|^2 + \nonumber\\
		&\quad \epsilon \tilde{\eta} \dfrac{\partial}{\partial y}\Bigg[|\nabla (\phi_b + \epsilon \tilde{\phi})|^2\Bigg]\Biggr) + g(\eta_b + \epsilon \tilde{\eta}) = 0 \label{eqc7}
	\end{align}
	We can eliminate the terms of $O(1)$, as they form the exact equations ((\ref{eq15})-(\ref{eq16})) satisfied by the base-state. Retaining only upto $O(\epsilon \hat{A}^2)$ and collecting the terms, yields the following \textcolor{black}{equation}:
	\begin{align}
		&\dfrac{\partial \tilde{\phi}}{\partial t} + \tilde{\eta}\dfrac{\partial^2 \phi_b}{\partial y \partial t} + \Biggl(\dfrac{\partial \phi_b}{\partial x}\dfrac{\partial \tilde{\phi}}{\partial x} + \dfrac{\partial \phi_b}{\partial y}\dfrac{\partial \tilde{\phi}}{\partial y}\Biggr) +  \dfrac{\tilde{\eta}}{2}\dfrac{\partial}{\partial y}\left(|\nabla \phi_b|^2\right) + g\tilde{\eta}=0 \label{eqc8}
	\end{align}
	Equation (\ref{eqc6}) and (\ref{eqc8}) are the coupled PDEs that govern the spatio-temporal evolution of quantities $\tilde{\eta}, \tilde{\phi}$, given the \textcolor{black}{form of} $\eta_b, \phi_b$ (base-state).




\end{appendices}


\bibliography{sn-bibliography}

@PREAMBLE{
	"\providecommand{\noopsort}[1]{}" 
	# "\providecommand{\singleletter}[1]{#1}%" 
}

@Book{Nayfeh,
	author       = {Nayfeh, A. H. and Mook, D. T},
	year         = 1995,
	title        = {Nonlinear Oscillations},
	publisher    = {John Wiley \& Sons.},
	address	     = "New York"
}

@Book{Stoker,
	author       = {Stoker, J. J.},
	year         = 1992,
	title        = {Nonlinear Vibrations in Mechanical and Electrical Systems},
	publisher    = {Wiley Classics Library, Reprint Edition},
	address	     = "New York"
}

@BOOK{Rand,
	author       = {Rand, R. H},
	year         = 2012,
	title        = {Lecture Notes on Nonlinear Vibrations},
	publisher    = {Available online: \url{https://ecommons.cornell.edu/handle/1813/28989}},
	address		= "Cornell University Library"
}

@ARTICLE{KovacicRand,
	author       = {Kovacic, I. and Rand, R. H. and Mohamed, Sah S.}, 
	year         = "2018", 
	journal      = " Applied Mechanics Reviews", 
	volume       = "70", 
	pages        = "020802-1--020802-22",
}

@ARTICLE{Penney1952,
	author       = "W. G. Penney and A. T. Price",
	title        = "FINITE PERIODIC STATIONARY GRAVITY WAVES IN A PERFECT LIQUID",
	journal      = "Philosophical Transactions of the Royal Society of London. Series A, Mathematical and Physical Sciences",
	volume       = "244", 
	pages        = "254–284",
	year         = "1952",
}

@book{ibrahim2005liquid,
	title={Liquid sloshing dynamics: theory and applications},
	author={Ibrahim, Raouf A},
	year={2005},
	publisher={Cambridge University Press},
	  address   = {Cambridge}
}

@book{bender2013advanced,
	title={Advanced mathematical methods for scientists and engineers I: Asymptotic methods and perturbation theory},
	author={Bender, Carl M and Orszag, Steven A},
	year={2013},
	publisher={Springer Science \& Business Media},
	  address   = {New York}
}

@article{molin2001piston,
	title={On the piston and sloshing modes in moonpools},
	author={Molin, Bernard},
	journal={Journal of Fluid Mechanics},
	volume={430},
	pages={27--50},
	year={2001},
	publisher={Cambridge University Press}
}

@article{miles2002gravity,
	title={Gravity waves in a circular well},
	author={Miles, John},
	journal={Journal of Fluid Mechanics},
	volume={460},
	pages={177--180},
	year={2002},
	publisher={Cambridge University Press}
}

@article{turner2024dynamic,
	title={Dynamic sloshing in a rectangular vessel with porous baffles},
	author={Turner, MR},
	journal={Journal of engineering mathematics},
	volume={144},
	number={1},
	pages={22},
	year={2024},
	publisher={Springer}
}

@article{chu2024nonlinear,
	title={On the nonlinear moonpool responses in a drillship under regular heading waves},
	author={Chu, Bei and Zhang, Xinshu and Zhang, Guangming and Chen, Junxuan},
	journal={Physics of Fluids},
	volume={36},
	number={3},
	year={2024},
	publisher={AIP Publishing}
}

@article{strutt1915deep,
	title={Deep water waves, progressive or stationary, to the third order of approximation},
	author={Strutt, John William},
	journal={Proceedings of the Royal Society of London. Series A, Containing Papers of a Mathematical and Physical Character},
	volume={91},
	number={629},
	pages={345--353},
	year={1915},
	publisher={The Royal Society London}
}

@misc{popinet-basilisk,
	author = {Popinet, S. and collaborators},
	howpublished = {\url{http://basilisk.fr} (Last accessed: August 23, 2023)},
	title = {Basilisk {C}: volume of fluid method},
	year = {2013--2024}}

@misc{Mathematica,
	author = {Wolfram Research{,} Inc.},
	title = {Mathematica, {V}ersion 12.1},
	url = {https://www.wolfram.com/mathematica},
	note = {Champaign, IL, 2024}
}

@misc{CoastalWiki,
	title        = {Harbour Resonance},
	author       = {Coastalwiki},
	year         = 2020,
	note         = {\url{https://www.coastalwiki.org/wiki/Harbor_resonance} [Accessed: May 14th 2025]}
}

@article{miles1974harbor,
	title={Harbor seiching},
	author={Miles, John W},
	journal={Annual Review of Fluid Mechanics},
	volume={6},
	number={1},
	pages={17--33},
	year={1974},
	publisher={Annual Reviews 4139 El Camino Way, PO Box 10139, Palo Alto, CA 94303-0139, USA}
}

@article{zhu2003three,
	title={Three-dimensional instability of standing waves},
	author={Zhu, Qiang and Liu, Yuming and Yue, Dick KP},
	journal={Journal of Fluid Mechanics},
	volume={496},
	pages={213--242},
	year={2003},
	publisher={Cambridge University Press}
}

@article{schwartz1981semi,
	title={A semi-analytic solution for nonlinear standing waves in deep water},
	author={Schwartz, LW and Whitney, AK},
	journal={Journal of Fluid Mechanics},
	volume={107},
	pages={147--171},
	year={1981},
	publisher={Cambridge University Press}
}

@article{rottman1982steep,
	title={Steep standing waves at a fluid interface},
	author={Rottman, James W},
	journal={Journal of Fluid Mechanics},
	volume={124},
	pages={283--306},
	year={1982},
	publisher={Cambridge University Press}
}

@article{mack1962periodic,
	title={Periodic, finite-amplitude, axisymmetric gravity waves},
	author={Mack, Lawrence R},
	journal={Journal of Geophysical Research},
	volume={67},
	number={2},
	pages={829--843},
	year={1962},
	publisher={Wiley Online Library}
}

@article{concus1962standing,
	title={Standing capillary-gravity waves of finite amplitude},
	author={Concus, Paul},
	journal={Journal of Fluid Mechanics},
	volume={14},
	number={4},
	pages={568--576},
	year={1962},
	publisher={Cambridge University Press}
}

@article{rycroft2013computation,
	title={Computation of three-dimensional standing water waves},
	author={Rycroft, Chris H and Wilkening, Jon},
	journal={Journal of Computational Physics},
	volume={255},
	pages={612--638},
	year={2013},
	publisher={Elsevier}
}

@article{saffman1979note,
	title={A note on numerical computations of large amplitude standing waves},
	author={Saffman, Philip G and Yuen, Henry C},
	journal={Journal of Fluid Mechanics},
	volume={95},
	number={4},
	pages={707--715},
	year={1979},
	publisher={Cambridge University Press}
}

@article{Mercer1992,
	author = {Mercer, G. N. and Roberts, A. J.},
	title = {Standing waves in deep water: Their stability and extreme form},
	journal = {Physics of Fluids A: Fluid Dynamics},
	volume = {4},
	number = {2},
	pages = {259-269},
	year = {1992},
	month = {02},
	issn = {0899-8213},
	doi = {10.1063/1.858354},
}

@article{YANG1967,
	title = {On the vibrations of a particle in the plane},
	journal = {International Journal of Non-Linear Mechanics},
	volume = {2},
	number = {1},
	pages = {1-25},
	year = {1967},
	issn = {0020-7462},
	doi = {https://doi.org/10.1016/0020-7462(67)90015-7},
	author = {Ta-Lun Yang and R.M. Rosenberg},
}

@article{YANG1968,
	title = {On forced vibrations of a particle in the plane},
	journal = {International Journal of Non-Linear Mechanics},
	volume = {3},
	number = {1},
	pages = {47-63},
	year = {1968},
	issn = {0020-7462},
	doi = {https://doi.org/10.1016/0020-7462(68)90024-3},
	author = {Ta-lun Yang and R.M. Rosenberg},
}

@article{PECELLI198057,
	title = {Stability of modes for a class of non-linear planar oscillators},
	journal = {International Journal of Non-Linear Mechanics},
	volume = {15},
	number = {1},
	pages = {57-70},
	year = {1980},
	issn = {0020-7462},
	doi = {https://doi.org/10.1016/0020-7462(80)90053-0},
	author = {G. Pecelli and E.S. Thomas},

}

@article{RandandFu,
	author = {Rand, Richard H. and Tseng, Shoei-Fu},
	title = {On the Stability of a Differential Equation With Application to the Vibrations of a Particle in the Plane},
	journal = {Journal of Applied Mechanics},
	volume = {36},
	number = {2},
	pages = {311-313},
	year = {1969},
	month = {06},
	issn = {0021-8936},
	doi = {10.1115/1.3564628},

}

@article{okamura1984instabilities,
	title={Instabilities of weakly nonlinear standing gravity waves},
	author={Okamura, Makoto},
	journal={Journal of the Physical Society of Japan},
	volume={53},
	number={11},
	pages={3788--3796},
	year={1984},
	publisher={The Physical Society of Japan}
}

@article{Taylor1953,
	author = {Taylor, Geoffrey Ingram },
	title = {An experimental study of standing waves},
	journal = {Proceedings of the Royal Society of London. Series A. Mathematical and Physical Sciences},
	volume = {218},
	number = {1132},
	pages = {44-59},
	year = {1953}
}

@article{rajchenbach2015faraday,
	title={Faraday waves: their dispersion relation, nature of bifurcation and wavenumber selection revisited},
	author={Rajchenbach, Jean and Clamond, Didier},
	journal={Journal of Fluid Mechanics},
	volume={777},
	pages={R2},
	year={2015},
	publisher={Cambridge University Press}
}

@article{fauve1998pattern,
	title={Pattern forming instabilities},
	author={Fauve, Stephan},
	journal={COLLECTION ALEA SACLAY MONOGRAPHS AND TEXTS IN STATISTICAL PHYSICS},
	pages={387--492},
	year={1998},
	publisher={Cambridge University Press}
}

@article{kayal2025focussing,
	title={Focussing of concentric free-surface waves},
	author={Kayal, Lohit and Sanjay, Vatsal and Yewale, Nikhil and Kumar, Anil and Dasgupta, Ratul},
	journal={Journal of Fluid Mechanics},
	volume={1003},
	pages={A14},
	year={2025},
	publisher={Cambridge University Press}
}

@article{basak2021jetting,
	title={Jetting in finite-amplitude, free, capillary-gravity waves},
	author={Basak, Saswata and Farsoiya, Palas Kumar and Dasgupta, Ratul},
	journal={Journal of Fluid Mechanics},
	volume={909},
	pages={A3},
	year={2021},
	publisher={Cambridge University Press}
}

@article{kayal2023jet,
	title={Jet from a very large, axisymmetric, surface-gravity wave},
	author={Kayal, Lohit and Dasgupta, Ratul},
	journal={Journal of Fluid Mechanics},
	volume={975},
	pages={A22},
	year={2023},
	publisher={Cambridge University Press}
}

@article{rackauckas2017differentialequations,
	author = {Rackauckas, Christopher and Nie, Qing and collaborators},
	journal = {Journal of Open Research Software},
	number = {1},
	pages = {15},
	publisher = {Ubiquity Press},
	title = {Differentialequations.jl--a performant and feature-rich ecosystem for solving differential equations in julia},
	volume = {5},
	year = {2017}}

@article{kayal2022dimples,
	title={Dimples, jets and self-similarity in nonlinear capillary waves},
	author={Kayal, Lohit and Basak, Saswata and Dasgupta, Ratul},
	journal={Journal of Fluid Mechanics},
	volume={951},
	pages={A26},
	year={2022},
	publisher={Cambridge University Press}
}

@inbook{Stokes_2009, place={Cambridge}, series={Cambridge Library Collection -
	Mathematics}, title={On the Theory of Oscillatory Waves}, 
	booktitle={Mathematical and Physical Papers}, 
	publisher={Cambridge University Press},
	volume = "1", 
	author={Stokes, George Gabriel}, 
	year={1880}, 
	pages={197–229},
	collection={Cambridge Library Collection - Mathematics},
	address = "Cambridge"
}

@article{Tsamopoulos_Brown_1983, title={Nonlinear oscillations of inviscid drops and bubbles}, volume={127}, DOI={10.1017/S0022112083002864}, journal={Journal of Fluid Mechanics}, author={Tsamopoulos, John A. and Brown, Robert A.}, year={1983}, pages={519–537}
}

@inbook{Kundu2016,
	title = {Chapter 8 - Gravity Waves},
	editor = {Pijush K. Kundu and Ira M. Cohen and David R. Dowling},
	booktitle = {Fluid Mechanics (Sixth Edition)},
	publisher = {Academic Press},
	edition = {Sixth Edition},
	address = {Boston},
	pages = {349-407},
	year = {2016},
	isbn = {978-0-12-405935-1},
	doi = {https://doi.org/10.1016/B978-0-12-405935-1.00008-3},
	author = {Pijush K. Kundu and Ira M. Cohen and David R. Dowling},
}

@book{magnus2013hill,
	title={Hill's equation},
	author={Magnus, Wilhelm and Winkler, Stanley},
	year={2013},
	publisher={Dover Publications },
	address	= "New York"
}

@article{LonguetHiggins1978,
	author = {Longuet-Higgins, Michael Selwyn },
	title = {The instabilities of gravity waves of finite amplitude in deep water II. Subharmonics},
	journal = {Proceedings of the Royal Society of London. A. Mathematical and Physical Sciences},
	volume = {360},
	number = {1703},
	pages = {489-505},
	year = {1978},
	doi = {10.1098/rspa.1978.0081}
}

@book{Cross_Greenside_2009, 
	place={Cambridge}, 
	title={Pattern Formation and Dynamics in Nonequilibrium Systems}, publisher={Cambridge University Press}, 
	author={Cross, Michael and Greenside, Henry},
	year={2009},
	address = "Cambridge"
}

@incollection{koszalka2005vibrating,
	title={Vibrating pendulum and stratified fluids},
	author={Koszalka, Inga Monika},
	booktitle={2005 Program in Geophysical Fluid Dynamics: Fast Times and Fine Scales},
	pages={205--224},
	year={2005},
	publisher={Woods Hole Oceanographic Institution},
	address = "Woods Hole, MA 02543 USA"
}

\end{document}